\documentclass{article}
\RequirePackage{fix-cm}

\usepackage{graphicx}
\usepackage{mathptmx}      
\DeclareMathAlphabet{\mathcal}{OMS}{cmsy}{m}{n}  
\DeclareSymbolFont{largesymbols}{OMX}{cmex}{m}{n} 

\usepackage{latexsym}
\usepackage{amsmath}
\usepackage{subfigure}
\usepackage{amssymb}
\usepackage{mathtools}
\usepackage{fullpage}
\usepackage[affil-it]{authblk}

\begin{document}

\title{Interdependency and hierarchy of exact and approximate epidemic models on networks}


\date{}
\author{Timothy J Taylor
\thanks{Timothy Taylor is funded by a PGR studentship 
from the MRC, and the Departments of Informatics and Mathematics at the University of Sussex.}}
\affil{              Centre for Computational Neuroscience and Robotics,\\ University of Sussex, Falmer, Brighton BN1 9QH, UK \\}

\author{Istvan Z Kiss
  \thanks{Corresponding Author. Electronic address: \texttt{I.Z.Kiss@sussex.ac.uk}}}
\affil{School of Mathematical and Physical Sciences, Department of Mathematics,\\ University of Sussex, Falmer, Brighton BN1 9QH, UK}

\maketitle

\begin{abstract}
Over the years numerous models of $SIS$ (susceptible $\rightarrow$ infected $\rightarrow$ susceptible) disease dynamics unfolding on networks have been proposed.
Here, we discuss the links between many of these models and how they can be viewed as more general motif-based models.
We illustrate how the different models can be derived from one another and, where this is not possible, discuss extensions 
to established models that enables this derivation.  We also derive a general result for the exact differential equations
for the expected number of an arbitrary motif directly from the Kolmogorov/master equations and conclude with a comparison of the performance of 
the different closed systems of equations on networks of varying structure.
\end{abstract}
\section{Introduction}
Modeling the spread of infectious diseases requires an understanding of not only disease characteristics but also
an understanding of the community (be it a hospital, school, town, etc) in which it pervades.  An important 
consideration in modelling the spread of diseases is thus the contact structure on which disease transmission happens.  
Whereas traditional approaches (\cite{anderson,diekmann}) assume little or no topological structure, 
recent work (\cite{keeling1,kenah,lindquist}) has tried to incorportate the underlying linkages between 
entities in the population and study how these links facilitate the spread of the disease.  For a continuous-time
stochastic disease transmission model on an arbitrary network it is possible (\cite{istvan1}),  to write down the relevant 
Kolmogorov/master equations and thus model it as a continuous time Markov chain that fully describes the movement between 
all possible system states.  Unfortunately the complexity of the model comes from the size of the state space and
the number of equations scales exponentially as $a^N$, where $a$ is the number of different states a node can be in 
and $N$ is the network size.  One widely used resolution to this complexity is to create individual-based simulation 
models and investigate the system behaviour directly.  Even though increasing computational power makes simulations an 
increasingly attractive proposition they lack analytic tractability.  Whilst this is not always a hindrance, when the 
system displays a rich range of behaviour (e.g.\ oscillations, bistability) it may not be feasable 
to obtain a global overview of the effects of different parameter values and thus the more analytic approach is needed.
For this reason, low-dimensional systems of differential equations (\cite{keeling1,eames1,lindquist}) are 
sought provided that these can approximate the exact solution.  By reducing the problem to a smaller system of equations 
it is easier to study the bifurcation structure of the model and gain a greater understanding of the full spectrum of behaviour.  
The challenge is then finding the set of equations that best approximate the solution of the Kolmogorov equations.

Given that here we focus on epidemic models, usually such models are formulated in terms of the expected values of the number 
of infected and/or susceptible individuals or some other motif in the network such as the expected number of 
infected and/or susceptible individuals of different degrees (the number of connections a node has). Such models range from
classic meanfield \cite{allen} to pairwise \cite{keeling1}, heterogenous pairwise \cite{eames1}, effective 
degree~\cite{lindquist,marceau} and individual-level models \cite{sharkey} to name a few.  Whilst these models 
seem to use different approaches their derivation is based on the same conceptual framework, namely they begin by choosing a base-motif 
(e.g.\ a node, a link and the two nodes it connects, a node and all its links).  These base-motifs are then used to 
formulate equations for the different possible states that they can achieve (e.g. for the expected number of motifs in different 
states or the probability that a specified motif in the network is in a certain state). These equations generally involve not only
the base-motif itself, but larger or extended motifs of which they are usually part of.  These larger motifs in turn depend on 
more complex motifs and a closure is needed in order to obtain a self-contained system of equations of reasonable size.
Importantly the base-motif determines not only the complexity of the model (the larger the motif the greater the number of states 
it can be in) but also how much of the network topology can be captured.  Interestingly 
differential equations for smaller motifs that are part of the base-motif should, in theory, be recoverable from the original 
differential equation.  To this end the main focus of the paper is the consideration of various simple models of disease dynamics 
and the relations between them.  We also consider which models are derivable directly (subject to a suitable closure) 
from the Kolmogorov/master equations and can thus be referred to as exact.  

We begin in section~\ref{sec:2} with an introduction of some of the more common approaches to modelling disease dynamics on 
networks, considering meanfield (\cite{allen}), pairwise (\cite{keeling1}), heterogeneous pairwise (\cite{eames1}) and the 
effective degree (\cite{lindquist}) model formulations.  In section~\ref{sec:3} we formulate an exact version of the effective 
degree model and then illustrate how the pairwise model can then be recovered from this new set of equations.  We are, however, 
unable to recover the heterogenous pairwise model from the exact effective degree and this motivates, in section~\ref{sec:4}, 
an extension of this which incorporates further network topology into the ODEs.  From this extension we then show how it is then 
possible to recover the heterogeneous pairwise equations.  Once the links between the models have been established, in
section~\ref{sec:5} we show how the unclosed version of the models can be derived directly from the Kolmogorov equations.
This is done by proving that as long as the heuristic equations for any motif are written following a certain set of rules they will always be exact. We 
conclude, in section~\ref{sec:6} with a brief comparison of the models and discuss under what circumstances they perform best,
in the sense of being close to simulation results.
\label{sec:1}

\section{Models of disease dynamics}
In this paper we focus on susceptible $\rightarrow$ infected $\rightarrow$ susceptible ($SIS$) disease dynamics on networks but 
note that all of the following models can be adapted for other disease (e.g. $SIR$ and/or contact tracing) or non-disease 
(e.g. evolutionary \cite{hadji}) dynamics.  With this in mind we use $\tau$ as the per-link transmission rate between 
susceptible and infected nodes and $\gamma$  as the recovery rate of an infected individual.  Both infection and recovery
are modelled as independent poisson processes.  As a starting point we give a short summary of ODE-based models that are either 
exact or an approximation of the true dynamics resulting from the full system based on the Kolmogorov/master equations, where these 
are solvable, or based on simulation.
\label{sec:2}

\subsection{Pairwise and the resulting simple compartmental model}
In order to focus on the underlying network of contacts, we introduce the pairwise model first (\cite{keeling1,rand1}).  
The main idea of this model is to develop the hierarchical dependence of lower order moments (e.g. expected number of susceptible 
$[S]$ and infected $[I]$ nodes) on higher ones (e.g. expected number of pairs with one susceptible and one infected node, $[SI]$) 
and to derive appropriate models that correctly account for these. As already suggested, the expected number of pairs will depend 
on larger motifs, in this case these being the expected number of triples denoted by $[ABC]$, where $A, B, C \in \{S, I\}$ and 
$B$ is connected to $A$ and $C$. Using this notation the equations governing the evolution of the disease dynamics 
at the level of singles and pairs are given by
\begin{align}
		\frac{d}{dt}\left[I\right] &= -\gamma\left[I\right] + \tau\left[S I\right],  \label{PairwiseI}\\
		\frac{d}{dt}\left[SS\right] &= -2\tau[ISS] + 2\gamma[IS], \label{PairwiseSS} \\
		\frac{d}{dt}\left[SI\right] &= \tau\left([ISS] - [ISI] - [IS]\right) + \gamma\left([II] - [IS]\right), \label{PairwiseSI}\\
		\frac{d}{dt}\left[II\right] &= 2\tau\left([ISI] + [SI]\right) - 2\gamma[II]. \label{PairwiseII}
\end{align}
Here we consider ordered pairs and triples, meaning they are counted in both directions.  For pairs (with similar remarks for 
triples) this means $[IS] = [SI]$ and links of type $S-S$ and $I-I$ have a double contribution to the $[SS]$ and $[II]$ counts.
Importantly we note that these equations are unclosed as no equations are given for the evolution of the triples.  The 
standard closure (in the absence of clustering) makes the assumption that the status of pairs are statistically
independent of one another and then 
\begin{align*}
\left[ABC\right] \approx &[AB](n-1)\frac{[BC]}{n[B]},
\end{align*}
where $n$ is the average degree of the network.  When we use this closure we say we have closed ``at the level of triples'' .
In order to derive the classic mean-field model a closure at the level of paris can be applied, namely, $[SI]$ can be approximated as
\begin{align*}
		[SI] \approx n[S]\frac{[I]}{N}
\end{align*}
and upon using Eq. (\ref{PairwiseI}), the classic mean-field model can be recovered
\begin{align}
		\frac{d}{dt}\left[I\right] &= -\gamma\left[I\right] + \tau n \left[S\right]\frac{\left[I\right]}{N},
\end{align}
where the widely used transmission rate from the compartmental model,\cite{allen}, is $\beta=\tau n$.

It is also important to note that the unclosed equations  above (Eqs. (\ref{PairwiseI}-\ref{PairwiseII})) can be derived
directly from the state-based Kolmogorov equations and for this reason we refer to these equations as exact.  
Whilst a proof for the exactness of these equations was given in~\cite{taylor1}, in section~\ref{sec:5} we provide
a more general proof that allows us to write down exact equations for, not just pairs, but any motif structure.  We
also note that an alternative approach was used by Sharkey in~\cite{sharkey}, to prove that the standard pairwise
equations were exact for models with susceptible $\rightarrow$ infected $\rightarrow$ recovered ($SIR$)  disease dynamics.
\label{sec:2a}

\subsection{Heterogeneous pairwise model}
Whilst the pairwise equations perform well in capturing disease dynamics on networks that are well described by their average 
degree, the closure assumption fails when greater heterogeneity is introduced.  More precisely, whilst the pairwise equations 
above are exact for an arbitrary network before a closure, these do not guarantee that with the current choice of singles and 
pairs (i.e. $[S]$ could be further divided to account for heterogeneity in degree) a valid closure could be found for any 
network. Indeed, to account for greater heterogeneity Eames \emph{et\ al.}~\cite{eames1} further developed the pairwise model by taking 
into account not just the state of nodes and pairs but also the degrees of the nodes.  By using $[A^n]$ to represent expected 
number of nodes of type $A$ with degree $n$ and with similar notation for pairs and triples, they were able to formulate the 
following set of unclosed equations
\begin{align}
\frac{d}{dt}\left[S^n\right] &= \gamma\left[I^n\right] - \tau\sum_{q}{\left[S^n I^q\right]}, \\
\frac{d}{dt}\left[I^n\right] &= -\gamma\left[I^n\right] + \tau\sum_{q}{\left[S^n I^q\right]},\\
\frac{d}{dt}\left[S^n S^m\right] &= -\tau\sum_{q}{\left(\left[S^n S^m I^q\right] + \left[I^q S^n S^m\right]\right)}
																	+ \gamma\left(\left[S^n I^m\right] + \left[I^n S^m\right]\right), \\
\frac{d}{dt}\left[S^n I^m\right] &= \tau\sum_{q}{\left(\left[S^n S^m I^q\right] - \left[I^q S^n I^m\right]\right)}
																	-\tau\left[S^n I^m\right] - \gamma\left[S^n I^m\right] + \gamma\left[I^n I^m\right], \\
\frac{d}{dt}\left[I^n I^m\right] &= \tau\sum_{q}{\left(\left[I^n S^m I^q\right] + \left[I^q S^n I^m\right]\right)}
																	+\tau\left[I^n S^m\right] + \tau\left[S^n I^m\right]  - 2\gamma\left[I^n I^m\right].
\end{align}
Again assuming the statistical independence of pairs and absence of clustering, Eames et.\ al,~\cite{eames1}, suggest the 
following approximations of triples
\begin{align*}
\left[B^nC^mD^p\right] \approx &[B^nC^m](m-1)\frac{[C^mD^p]}{m[C^m]}.
\end{align*}
\label{sec:2b}

\subsection{The effective degree model}
In~\cite{lindquist}, Lindquist \emph{et\  al.} introduced the effective degree model for $SIS$ (and also $SIR$) dynamics on a network
(an equivalent model formulation was also proposed by Marceau et al. \cite{marceau}).  In this model they consider not only the state
of a node ($S$ or $I$), but also the number of the immediate neighbours in the various potential states.  This is done by writing the 
following set of equations for all the possible star-like motifs in the network where $S_{s,i}$ ($I_{s,i}$) represents the 
expected number of susceptible (infected) nodes with $s$ susceptible and $i$ infected neighbours,
\begin{align}
\dot{S}_{s,i}=&-\tau i S_{s,i} + \gamma I_{s,i} +\gamma [(i+1)S_{s-1,i+1}-iS_{s,i}] \notag \\
&+ \tau \frac{\sum_{k=1}^{M}\sum_{j+l=k}jlS_{j,l}}{\sum_{k=1}^{M}\sum_{j+l=k}jS_{j,l}}[(s+1)S_{s+1,i-1} -sS_{s,i}], \\
\notag \\
\dot{I}_{s,i}=&\tau i S_{s,i} - \gamma I_{s,i} +\gamma [(i+1)I_{s-1,i+1}-iI_{s,i}] \notag \\
&+ \tau \frac{\sum_{k=1}^{M}\sum_{j+l=k}l^{2}S_{j,l}}{\sum_{k=1}^{M}\sum_{j+l=k}jI_{j,l}}[(s+1)I_{s+1,i-1} -sI_{s,i}],
\end{align}
with $1 \leq s+i \leq M$, where $M$ is the maximum degree and the equations are suitably adjusted on the boundaries. 
It is important to note that this model is not exact as a closure has been already applied.  Namely the infection of 
a node's susceptible neighbours is based on a population-level approximation.  To illustrate this more precisely we 
borrow the notation of the pairwise model and make two observations
\begin{align*}
\frac{\sum_{k=1}^{M}\sum_{j+l=k}jlS_{j,l}}{\sum_{k=1}^{M}\sum_{j+l=k}jS_{j,l}}
		&= \frac{[ISS]}{[SS]}, \\
\frac{\sum_{k=1}^{M}\sum_{j+l=k}l^{2}S_{j,l}}{\sum_{k=1}^{M}\sum_{j+l=k}jI_{j,l}}
		&= \frac{\sum_{k=1}^{M}\sum_{j+l=k}l(l-1)S_{j,l} + lS_{j,l}}{\sum_{k=1}^{M}\sum_{j+l=k}jI_{j,l}} = \frac{[ISI] + [SI]}{[SI]} = \frac{[ISI]}{[SI]} + 1.
\end{align*}
These means that the infection pressure on the susceptible neighbours of the central node is equal to the population level 
average taken from all the possible star-like configurations rather then from the extended star structures that would account 
exactly for these infections.
\label{sec:2c}
\section{Recovering the pairwise model from the effective degree}
Whilst the pairwise and effective degree models seem different they are based on a similar approach. Both models 
work on approximating the evolution of different motifs in the network; individuals and links in the pairwise model 
and star-like structures in the effective degree.  For both models, but more clearly for the pairwise, the models 
begin with a starting or base motif (e.g. nodes) for which an evolution equation is required. This will of course 
depend on an extended motif, typically the base motif extended by the addition of an extra node (e.g. pairs). This 
dependency on higher order motifs continues, for example, with pairs depending on triples, and then triples 
depending on quadruplets (four nodes connected by a line, i.e. $A-B-C-D$, or a star with a centre and three spokes, 
i.e. $A-\overbracket[0.5pt]{B-C \quad D}$). Hence, the models only differ in the choice of the base motif and 
then potentially in the way in which the systems are closed to curtail the dependency on higher order motifs. 
Since, here we are mainly interested in exact models, that is before a closure is applied, we begin by conjecturing 
an exact version of the effective degree model and show how starting from this the exact pairwise model can be derived.
\label{sec:3}

\subsection{Exact effective degree}
\begin{figure}[htp!]
  	\begin{center}
    		\includegraphics[width=0.75\textwidth]{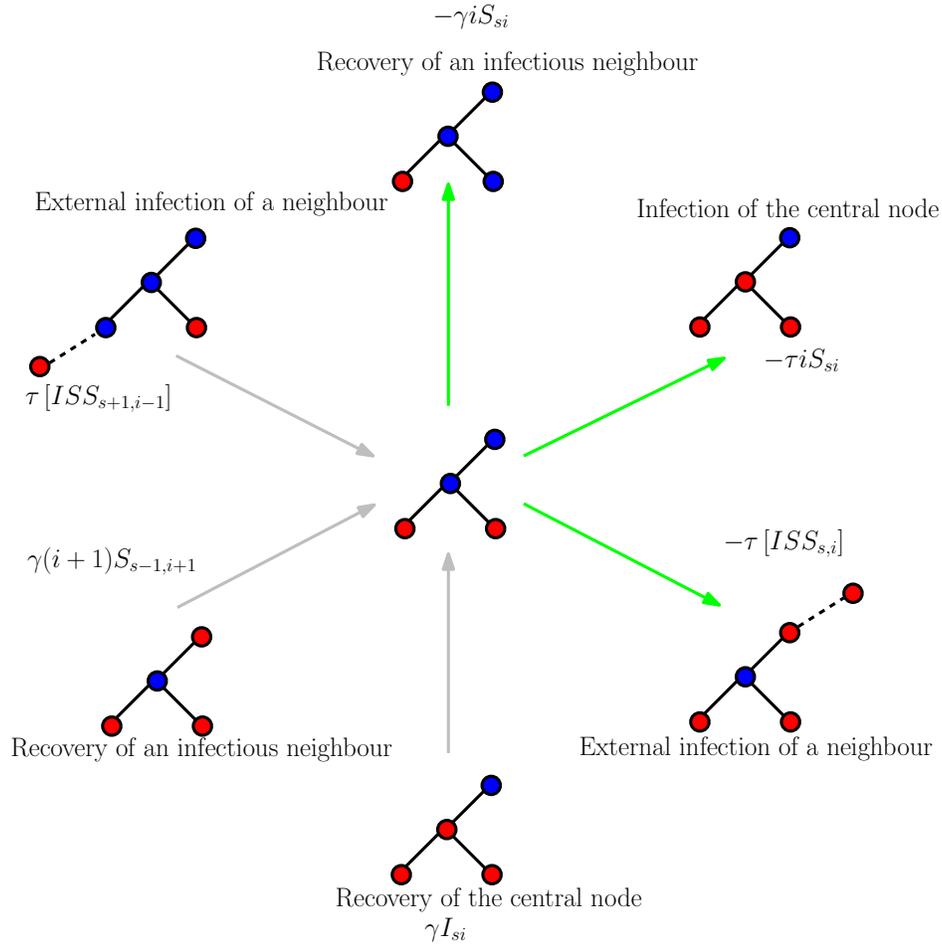}
  	\end{center}
  	\caption{Illustration of the transitions into and out of the $S_{2,1}$ class.  Susceptible nodes are given in blue
  		       	   and infective nodes in red. Transitions into and out of the class are shown in grey and green, respectively.
  				   The corresponding terms of the general equation are also given.  In Appendix~$1$ a similar illustration is
						given for a configuration with a centrally infectious node.}  
	\label{fig1}
\end{figure}

Based on the ideas presented above, we extend the star-like base motif to reveal the dependence on higher order motifs and conjecture that this unclosed version of the effective degree model is exact.  We begin by introducing a variable to count the expected number of infecteds connected to a node's susceptible neighbours.  This is done by introducing two new terms, $[ISS_{s',i'}]$ and 
$[ISI_{s',i'}]$.  For the term $[ISI_{s',i'}]$ (and similarly for $[ISS_{s',i'}]$) the $S$ in the middle is actually used to represent the susceptible neighbours of the central $I$  from the motif with composition $I_{s'i'}$ (i.e. the $I$ node with neighbourhood $(s',i')$ is the centre of the star, while $S$ is a susceptible spoke).  The $I$ (on the left-hand side), in turn, represents the infective neighbours of these susceptibles' and within this count, in the case of $[ISI_{s',i'}]$, we also include the originating central $I$.  The exact effective degree model can then be written as 
\begin{align}
\dot{S}_{s,i}=&-\tau i S_{s,i} + \gamma I_{s,i} +\gamma [(i+1)S_{s-1,i+1}-iS_{s,i}] \notag \\
&+ \tau\left[ISS_{s+1,i-1}\right] - \tau\left[ISS_{s,i}\right], \\ 
\notag \\
\dot{I}_{s,i}=&\tau i S_{s,i} - \gamma I_{s,i} +\gamma [(i+1)I_{s-1,i+1}-iI_{s,i}]\notag \\
&+ \tau\left[ISI_{s+1,i-1}\right] - \tau\left[ISI_{s,i}\right]. 
\end{align}
Fig.~\ref{fig1} shows the possible transitions captured by this model. 
\label{sec:3a}

\subsection{Recovering the pairwise equations}
The star-like composition of the effective degree model allows us to recover the pairwise equations via careful summations.  
The full derivation of the pairwise model is given in Appendix~$2$, whilst here we only illustrate the derivation of the
individuals (trivial but given for completeness) and the [$II$] pairs,

\begin{align*}
		\frac{d}{dt}\left[S\right] &= \sum_{s,i}\dot{S}_{s,i} = \gamma\left[I\right] - \tau\left[S I\right], \\
		\frac{d}{dt}\left[I\right] &= \sum_{s,i}\dot{I}_{s,i} = -\gamma\left[I\right] + \tau\left[S I\right],
\end{align*}
where most terms from the original effective degree equations cancel and we have used that $\sum_{s,i}iS_{s,i} = [SI]$ and
$\sum_{s,i}I_{s,i} = [I]$.  For $[II]$ the following equality holds

\begin{align*}
\frac{d}{dt}\left[II\right] 
			= &\sum_{s,i}i\dot{I}_{s,i} \\
			= &\tau\sum{i^2S_{s,i}} - \gamma\sum{iI_{s,i}}
				 +\gamma\sum{i(i+1)I_{s-1,i+1}} -\gamma\sum{i^2I_{s,i}} \\
				&+\tau\sum{i[ISI_{s+1,i-1}]} - \tau\sum{i[ISI_{s,i}]} \\
			= &\tau\sum{i(i-1)S_{s,i}} + \tau\sum{iS_{s,i}} - \gamma[II] \\
				&+\gamma[III] - \gamma\sum{i(i-1)I_{s,i}} -\gamma\sum{iI_{s,i}} \\
				&+\tau\sum{(i-1)[ISI_{s+1,i-1}]} + \tau\sum{[ISI_{s+1,i-1}]} - \tau\sum{i[ISI_{s,i}]} \\
			= &\tau[ISI] + \tau[IS] - \gamma[II] + \gamma[III] -\gamma[III] - \gamma[II] + \tau[ISI] + \tau[IS] \\
			= &2\tau\left([ISI] + [IS]\right) - 2\gamma[II],
\end{align*}
where we have used that $\sum_{s,i}iI_{si}=[II]$, $\sum_{s,i}{(i-1)[ISI_{s+1,i-1}]}=\sum{i[ISI_{s,i}]}$ and that $\sum{[ISI_{s+1,i-1}]}=[ISI]+[SI]$. 
These all follow from the definition of the pairwise model and the definition of the new extended motifs 
from the exact effective degree model.  We note that this result does indeed correspond to that of the 
given pairwise model.
\label{sec:3b}

\section{Higher order models}
Whilst we can recover the pairwise equations from the exact effective degree model we note that the same is not possible
with the heterogeneous pairwise equations.  This motivates an extension of the effective degree model where the 
degrees of neighbouring nodes are also taken in to account.  Again we conjecture that this model can, in theory, be
derived from the exact Kolmogorov equations and thus refer to it as exact.
\label{sec:4}

\subsection{Exact effective degree with neighbourhood composition}
We extend the exact effective degree model to include the number of neighbours of the central 
nodes' neighbours.  We begin by defining the following notation
\begin{align*}
s' &= (s_1,s_2, \ldots,s_M), \\
i' &= (i_1,i_2, \ldots,i_M), \\
|s'| &= s_1 + s_2 + \ldots + s_M, \\
|i'| &= i_1 + i_2 + \ldots + i_M, 
\end{align*}
where $s_{j}$ ($i_j$) represents the number of susceptible (infective) neighbours of degree $j$.  We now define $S_{s'i'}$, 
($I_{s'i'}$) as the number of susceptible (infective) nodes with neighbouring nodes whose own degrees are given by the 
entries in $s'$ and $i'$.  We can now write the extended ODEs in the following form
\begin{align}
\dot{S}_{s,'i'}= &-\tau |i'|S_{s,'i'} + \gamma I_{s',i'} 
														+\gamma \sum_{k=1}^{M}{(i_k+1)S_{s'_{k-},i'_{k+}}} -\gamma |i'|S_{s',i'} \notag \\
													&+\tau\sum_{k=1}^M{\left[IS^kS_{s'_{k+},i'_{k-}}\right]} -\tau\left[ISS_{s',i'}\right], \\
\dot{I}_{s',i'}= &\tau |i'|S_{s',i'} - \gamma I_{s',i'} 
														+\gamma \sum_{k=1}^{M}{(i_k+1)I_{s'_{k-},i'_{k+}}} -\gamma |i'|I_{s',i'} \notag \\
													&+\tau\sum_{k=1}^M{\left[IS^kI_{s'_{k+},i'_{k-}}\right]} -\tau\left[ISI_{s',i'}\right].													
\end{align}
Here $s'_{k-} = (s_1,s_2, \ldots, s_k -1, \ldots, s_M)$ and $s'_{k+} = (s_1,s_2, \ldots, s_k +1, \ldots, s_M)$ with a similar definition
for $i'_{k-}$ and $i'_{k+}$.  With a small modification to the exact effective degree notation terms such as $\left[IS^kS_{s'_{k+},i'_{k-}}\right]$ 
are taken to represent number of infectious contacts of the susceptible neighbours of degree $k$.
\label{sec:4a}

\subsection{Model recovery}
\begin{figure}[htp!]
  \begin{center}
    \includegraphics[width=0.75\textwidth]{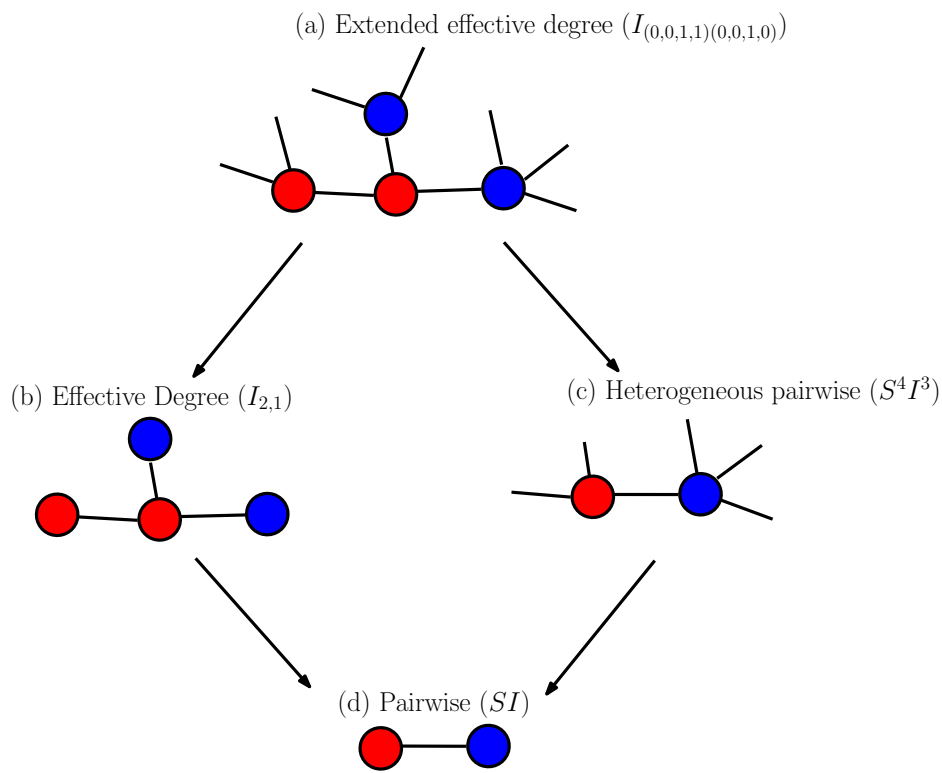}
  \end{center}
    \caption{Illustration of the hierarchial structure of model recovery.  Links that are known are given by lines
    and knowledge of a nodes status is given by circles.  Susceptible and infective nodes are shown in blue and red
		respectively. The upper level (a) represents the extended effective 
    degree ODEs.  The status of the central node is known along with that of it's neighbours and also their degrees.
    The secondary level is given by (b), the effective degree model where there is no knowledge of neighbours' 
    degrees and (c), the heterogenous pairwise model where the number of pairs of nodes and their relative degree is know.  
    The final level  shown, (d), is known as the standard pairwise model, \cite{keeling1}, where the status of individual nodes 
    and pairs is used.}
  \label{fig2}
\end{figure}

Here we show how, from the extended effective degree model, we can recover the heterogenous pairwise model. It is also 
straightforward to show, and thus omitted here, that the extended effective degree leads to the simpler exact effective degree.  
In turn, it also follows easily that both the exact effective degree and heterogenous pairwise models reduce to the  
standard pairwise model.  This hierarchy of recovery is illustrated in Fig.~\ref{fig2}.
\label{sec:4b}

\subsubsection{Recovering the heterogeneous pairwise model from the extended effective degree}
As earlier we make use of careful summation to recover the model.  The full derivation is provided in Appendix~$3$ so here
we just provide the derivation at the individual level and of the [$I^lI^n$] pairs.  For singles the following identities hold,
\begin{align*}
		\frac{d}{dt}\left[S^n\right] &= \sum_{|s'|+|i'|=n}\dot{S}_{s',i'} = \gamma\left[I^n\right] - \tau\left[S^n I\right], \\
		\frac{d}{dt}\left[I^n\right] &= \sum_{|s'|+|i'|=n}\dot{I}_{s',i'} =  -\gamma\left[I^n\right] + \tau\left[S^n I\right],
\end{align*}
where most terms from the original effective degree cancel and we have used that 
\begin{align*}
&\sum_{|s'|+|i'|=n}I_{s',i'} = [I^n] \quad \text{ and } \sum_{|s'|+|i'|=n}|i'|S_{s',i'} = [S^nI].
\end{align*}
For the $[I^lI^n]$ pair we obtain
\begin{align*}
\frac{d}{dt}\left[I^lI^n\right] 
			= &\sum_{|s'|+|i'|=n}{i_l\dot{I}_{s',i'}} \\
			= &\tau\sum{i_l|i'|S_{s',i'}} - \gamma\sum{i_lI_{s',i'}}
				 +\gamma\sum{i_l\sum_{k=1}^{M}{(i_{k}+1)I_{s'_{k-}i'_{k+}}}} \\
			  &-\gamma\sum{i_l|i'|I_{s',i'}} + \tau\sum{i_l\sum_{k=1}^{M}{\left[IS^kI_{s'_{k+},i'_{k-}}\right]}}
			   - \tau\sum{i_{l}\left[ISI_{s',i'}\right]} \\
			= &\tau\sum{i_l\left(|i'|-1\right)S_{s',i'}} + \tau\sum{i_lS_{s',i'}} - \gamma\left[I^lI^n\right] \\
			  &+\gamma\left[I^lI^nI\right] -\gamma\sum{i_l\left(|i'|-1\right)I_{s',i'}} - \gamma\sum{i'_lI_{s',i'}} \\
			  &+ \tau\sum{i_l\sum_{k \neq l}{\left[IS^kI_{s'_{k+},i'_{k-}}\right]}}
			  + \tau\sum{\left(i_l-1\right)\left[IS^lI_{s'_{l+},i'_{l-}}\right]}\\
			  &+ \tau\sum{\left[IS^lI_{s'_{l+},i'_{l-}}\right]} - \tau\sum{i_{l}\left[ISI_{s',i'}\right]}\\
			= &\tau\left[I^lS^nI\right] + \tau\left[I^lS^n\right] - 2\gamma\left[I^lI^n\right] \\
			  &+\gamma\left[I^lI^nI\right] - \gamma\left[I^lI^nI\right] + \tau\left[IS^lI^n\right] + \tau\left[S^lI^n\right] \\
			= &\tau\left[I^lS^nI\right] + \tau\left[I^lS^n\right] - 2\gamma\left[I^lI^n\right] + \tau\left[IS^lI^n\right] + \tau\left[S^lI^n\right] \\
			=&\tau\sum_{q}{\left(\left[I^l S^n I^q\right] + \left[I^q S^l I^n\right]\right)}
																	+\tau\left[I^l S^n\right] + \tau\left[S^l I^n\right]  - 2\gamma\left[I^l I^n\right].
\end{align*}
Again, we note that this result corresponds to previously given heterogenous pairwise model.

\label{sec:4b1}
\section{Exactness of the models}
In the previous sections we have at times referred to a set of ODEs as being exact.  This terminology implies that the 
ODEs can be derived directly from the Kolmogorov equations which describe the evolution of the epidemic through the 
full state space $\mathcal{S}$ (on a network of size $N$, $\mathcal{S}=\{S,I\}^N$).  In \cite{taylor1} the  exactness of 
the pairwise equations was rigorously proven but no other motif structures were considered.  In section~\ref{sec:3a}, 
we conjectured that the newly defined exact effective degree model is derivable from the
Kolmogorov equations.  Due to the structure of the motifs used in the effective degree model a mechanistic proof 
(as in~\cite{taylor1}) may be difficult and intricate to implement.  Instead we will prove that a heuristic 
formulation of the ODEs for any motif structure is indeed exact providing they are written following rigorous bookkeeping.
This derivation of the evolution equations for an arbitrary motif, directly from the Kolmogorv equations, will be based 
on an extension of ideas presented in \cite{istvan1} and \cite{taylor1} and using the notation defined in 
Tables~\ref{table:1}~and~{\ref{table:2}}.

We should note that in what follows a motif of connected nodes will only ever be counted once.  In a network of size
$N$ and considering a motif, $m$, with $k$ nodes this singular counting can be understood in the following way.  We consider each of the 
$\binom{N}{k}$ unique sets of $k$ nodes between $1$ and $N$.  Then for each set whose nodes are isomorphic in
topological structure and status to the motif $m$, we simply increase the counter of such motifs by one.  This formalism is unlike
that used in the standard pairwise model where an $SS$ link would contribute a value of two to the $[SS]$ count.  However,
the two resultant sets of equations are equivalent in the sense that the different ways of counting can easily be recovered 
by using a simple mapping between the two.  For this reason, whilst we prove that the following theorem is correct, it's intricacy 
and generality means a certain amount of care is needed when interpreting the resultant terms.  Using the notation defined
in Table~{\ref{table:2}} the result for a general motif is then given in the following theorem.
\label{sec:5}

\subsection*{Theorem 1}
The equation for the expected number ($|\mathcal{M}|$) of motifs of type $\hat{m}$, given by
\begin{align}
		\dot{|\mathcal{M}|} = 	&\tau\mathcal{N}_{in}^{SI}(\hat{m}^-,\hat{m}) + \tau\mathcal{N}_{ex}^{SI}(\hat{m}^-,\hat{m}) -\tau|\mathcal{M}|N_{in}^{SI}(\hat{m}) - \tau\mathcal{N}_{ex}^{SI}(\hat{m}) \notag \\
					&+\gamma\mathcal{N}^{I}(\hat{m}^+,\hat{m}) - \gamma|\mathcal{M}|N^{I}(\hat{m}) \label{heuristic}	
\end{align}
is derivable directly from the exact Kolmogorov equations.

\begin{table}
	\caption{Notation for matrix representation of the Kolmogorov equations (Table from~\cite{taylor1}).}
	\label{table:1}
	\begin{tabular}{ll}
		\hline\noalign{\smallskip}
		Variable & Definition  \\ \hline
		\noalign{\smallskip}\hline\noalign{\smallskip}
		$N$											   	& Number of nodes in the network \\ \noalign{\smallskip}\hline\noalign{\smallskip}
		$G=(g_{ij}) \in \{0,1\}^{N^2}$, $i,j=1,2,\ldots,N$ 	& \begin{minipage}[t]{0.55\columnwidth}
													      	   Adjacency matrix with $g_{ij}=1$ if nodes $i$ and
															   $j$ are connected and $g_{ij}=0$ otherwise.  The 
															   network is bi-directional and has no self loops such
															   that $G = G^T$ and $G_{ii} = 0$, $\forall$ $i$.
															   \end{minipage} \\ \noalign{\smallskip}\hline\noalign{\smallskip}
		$\tau$											   	& Rate of infection per ($S,I$) edge. \\ \noalign{\smallskip}\hline\noalign{\smallskip}
		$\gamma$											& Rate of recovery. \\ \noalign{\smallskip}\hline\noalign{\smallskip}
		$\mathcal{S} = \{S,I\}^N$							& \begin{minipage}[t]{0.55\columnwidth}
															    State space of the network, with nodes either
															    susceptible, $S$, or infected, $I$ and $|S|$ = $2^N$.
															    \end{minipage} \\ \noalign{\smallskip}\hline\noalign{\smallskip}
		$\mathcal{S}^k = \{\mathcal{S}_1^k,\mathcal{S}_2^k, \ldots, \mathcal{S}_{c_k}^k\}$
															& \begin{minipage}[t]{0.55\columnwidth}
															   The $c_k = \binom{N}{k}$ states with $k$ infected
															   individuals in all possible configurations, with 
															   $k = 0,1,\ldots,N$.
															   \end{minipage} \\ \noalign{\smallskip}\hline\noalign{\smallskip}
		$X_j^k(t)$											& \begin{minipage}[t]{0.55\columnwidth}
															    Probability of being in state $\mathcal{S}_j^k$ at time
															    $t$, where $k=0,1,\ldots,N$ and $j = 1,2,\ldots, c_k$.
															    \end{minipage} \\ \noalign{\smallskip}\hline\noalign{\smallskip}
		$X^k(t)$											& $X^k(t) = \left(X_1^k(t),X_2^k(t),\ldots, X_{c_k}^k(t)\right)^{T}$. 
															    \\ \noalign{\smallskip}\hline\noalign{\smallskip}															    
		$A^k_{i,j}$											& \begin{minipage}[t]{0.55\columnwidth}
															    Rate of transition from $\mathcal{S}_j^{k-1}$ to
															    $\mathcal{S}_i^{k}$, where $k = 0,1,\ldots, N$,
															    $i = 1,2,\ldots,c_k$ and $j = 1,2,\ldots,c_{k-1}$ .  Note that only one individual
															    is changing (i.e.\ in this case an $S$ node changes
															    to an $I$ through infection).
															    \end{minipage} \\ \noalign{\smallskip}\hline\noalign{\smallskip}
		$C^k_{i,j}$											& \begin{minipage}[t]{0.55\columnwidth}
															    Rate of transition from $\mathcal{S}_j^{k+1}$ to
															    $\mathcal{S}_i^{k}$, where $k = 0,1,\ldots, N$,
															    $i =1,2,\ldots,c_k$ and $j = 1,2,\ldots,c_{k+1}$.  
															    Note that only one individual is changing (i.e.\ in this case 
															    an $I$ node changes to an $S$ through recovery).
															    \end{minipage} \\ \noalign{\smallskip}\hline\noalign{\smallskip}
		$B^k_{i,j}$											& \begin{minipage}[t]{0.55\columnwidth}
															    Rate of transition from $\mathcal{S}_j^{k}$ to 
															    $\mathcal{S}_i^{k}$, where 
															    $B_{i,j}^k = 0$ if $i \neq j$ with $k = 0,1,\ldots, N$ 
															    and $i,j = 1,2,\ldots,c_k$.
															    \end{minipage} \\ \noalign{\smallskip}\hline\noalign{\smallskip}
		\noalign{\smallskip}\hline
	\end{tabular}
\end{table}
															    
\begin{table}
	\caption{Additional notation for matrix representation of the Kolmogorov equations}
	\label{table:2}
	\begin{tabular}{ll}
		\hline\noalign{\smallskip}
		Variable & Definition  \\ \hline
		\noalign{\smallskip}\hline\noalign{\smallskip}
		$\hat{m}$												& \begin{minipage}[t]{0.7\columnwidth}
															     An arbitrary motif encompassing both topology and status of nodes
															     (e.g.\ an $S-I$ edge or a star like structure such as $I_{3,0}$).
															    The arbitrary motif we are consdering which will encompass
															    both topology and status of nodes.
															    \end{minipage} \\ \noalign{\smallskip}\hline\noalign{\smallskip}															   
		$\hat{m}^+$												& \begin{minipage}[t]{0.7\columnwidth}
															    Represents the different motifs with the same structure as $\hat{m}$
															    but with a susceptible node of $\hat{m}$ having become infected.
															    \end{minipage} \\ \noalign{\smallskip}\hline\noalign{\smallskip}															    															 
		$\hat{m}^-$												& \begin{minipage}[t]{0.7\columnwidth}
															    Represents the different motifs with the same structure as $\hat{m}$
															    but with with an infective node of $\hat{m}$ having become susceptible.
															    \end{minipage} \\ \noalign{\smallskip}\hline\noalign{\smallskip}															    															 	
		$M_{k,j}$														& \begin{minipage}[t]{0.7\columnwidth}
															    Set of $\hat{m}$ motifs in configuration state $\mathcal{S}_j^k$.  
															    Defining the $i^{th}$ element of $M_{k,j}$ as $\hat{m}_{k,j}^i$
															    gives $M_{k,j} = \{ \hat{m}_{k,j}^1, \hat{m}_{k,j}^2, \ldots, \hat{m}_{k,j}^{|M|} \}$.
															    \end{minipage} \\ \noalign{\smallskip}\hline\noalign{\smallskip}															    
		$M_{k,j}^+$														& \begin{minipage}[t]{0.7\columnwidth}
															    The set of motifs, in configuration state $\mathcal{S}_j^k$, with the same topology as $\hat{m}$ but with $1$ more 
															    infective and $1$ less susceptible. Defining the $i^{th}$ element of $M_{k,j}^+$ as 
															    $\hat{m}_{k,j}^{i+}$ gives $M_{k,j}^{+} = \{ \hat{m}_{k,j}^{1+}, \hat{m}_{k,j}^{2+}, \ldots, \hat{m}_{k,j}^{|M_{k,j}^+|+} \}$.
															    \end{minipage} \\ \noalign{\smallskip}\hline\noalign{\smallskip}
		$M_{k,j}^-$														& \begin{minipage}[t]{0.7\columnwidth}
															    The set of motifs, in configuration state $\mathcal{S}_j^k$, with the same topology as $\hat{m}$ but with $1$ less 
															    infective and $1$ more susceptible.  Defining the $i^{th}$ element of $M_{k,j}^-$ as 
															    $\hat{m}_{k,j}^{i-}$ we have $M_{k,j}^- = \{ \hat{m}_{k,j}^{1-}, \hat{m}_{k,j}^{2-}, \ldots, \hat{m}_{k,j}^{|M_{k,j}^-|-} \}$.
															    \end{minipage} \\ \noalign{\smallskip}\hline\noalign{\smallskip}
		$N_{\hat{m}}(\mathcal{S}_j^k)$							& \begin{minipage}[t]{0.7\columnwidth}
															    Number of $\hat{m}$ motifs in state $\mathcal{S}_j^{k}$,
															    with $k = 0,1,\ldots, N$ and $j = 1,2,\ldots,c_k$.  
															    \end{minipage} \\ \noalign{\smallskip}\hline\noalign{\smallskip}
		$N_{in}^{SI}(\hat{h})$											& \begin{minipage}[t]{0.7\columnwidth}
															    Number of $SI$ links within the motif $\hat{h}$.
															    \end{minipage} \\ \noalign{\smallskip}\hline\noalign{\smallskip}
		$\mathcal{N}_{in}^{SI}(\hat{h})$						& \begin{minipage}[t]{0.7\columnwidth}
															    Expected total number of $SI$ links within all motifs of type $\hat{h}$
															    \end{minipage} \\ \noalign{\smallskip}\hline\noalign{\smallskip}
		$N_{in}^{SI}(\hat{h},k)$											& \begin{minipage}[t]{0.7\columnwidth}
															    Number of $SI$ links within the motif $\hat{h}$, along which, were an infection
															    to occur, would result in a motif of type $k$.
															    \end{minipage} \\ \noalign{\smallskip}\hline\noalign{\smallskip}
		$\mathcal{N}_{in}^{SI}(\hat{h},k)$						& \begin{minipage}[t]{0.7\columnwidth}
															    Expected total number of $SI$ links within all motifs of type $\hat{h}$, 
															    along which, were an infection to occur, would result in a motif of type $k$.
															    \end{minipage} \\ \noalign{\smallskip}\hline\noalign{\smallskip}															    								
		$N_{ex}^{SI}(\hat{h})$											& \begin{minipage}[t]{0.7\columnwidth}
															    Number of $SI$ links where the $S$ is contained within the motif $\hat{h}$ and the $I$ is external to it.
															    \end{minipage} \\ \noalign{\smallskip}\hline\noalign{\smallskip}
		$\mathcal{N}_{ex}^{SI}(\hat{h})$									& \begin{minipage}[t]{0.7\columnwidth}
															    Expected total number of $SI$ links to all motifs with structure $\hat{h}$, where the $S$ is contained within
															    the motif $\hat{h}$ and the $I$ external to it.
															    \end{minipage} \\ \noalign{\smallskip}\hline\noalign{\smallskip}
		$N_{ex}^{SI}(\hat{h},k)$											& \begin{minipage}[t]{0.7\columnwidth}
															    Number of $SI$ links where the $S$ is contained within the motif $\hat{h}$ and the $I$ is external to it,
															    along which, were an infection to occur, would result in a motif of type $k$.
															    \end{minipage} \\ \noalign{\smallskip}\hline\noalign{\smallskip}
		$\mathcal{N}_{ex}^{SI}(\hat{h},k)$								& \begin{minipage}[t]{0.7\columnwidth}
															    Expected total number of $SI$ links to all motifs with structure $\hat{h}$, where the $S$ is contained within
															    the motif $\hat{h}$ and the $I$ external to it, along which, were an infection to occur, would result in a 
															    motif of type $k$.
															    \end{minipage} \\ \noalign{\smallskip}\hline\noalign{\smallskip}
		$N^{I}(\hat{h})$													& \begin{minipage}[t]{0.7\columnwidth}
															    Number of $I$ nodes within motif $\hat{h}$.
															    \end{minipage} \\ \noalign{\smallskip}\hline\noalign{\smallskip}
		$N^{I}(\hat{h},k)$													& \begin{minipage}[t]{0.7\columnwidth}
															    Number of $I$ nodes within motif $\hat{h}$, whose recovery lead to a motif of type $k$.
															    \end{minipage} \\ \noalign{\smallskip}\hline\noalign{\smallskip}
		$\mathcal{N}^{I}(\hat{h},k)$											& \begin{minipage}[t]{0.7\columnwidth}
															    Expected total number of $I$s within motifs of type $\hat{h}$, whose recovery lead to a motif of type $k$.
															    \end{minipage} \\ \noalign{\smallskip}\hline\noalign{\smallskip}															    															  
		\noalign{\smallskip}\hline
	\end{tabular}
\end{table}

\subsection{Proof of Theorem 1}
For a detailed description of writing the Kolmogorov equations for an arbitrary graph we refer the reader
to \cite{istvan1}.  Here we only provide a brief description making use of the notation defined in 
Table~\ref{table:1}. 
The $2^N$ elements of the state space, $\mathcal{S}=\{S,I\}^N$, can be divided into $N+1$ subsets where,
for $0 \leq k \leq N, \mathcal{S}^k$ is the subset of all states with $k$ infected nodes.  Necessarily each subset
contains $c_k=\binom{N}{k}$ distinct configurations, i.e.\  $\mathcal{S}^k = (\mathcal{S}^k_1,\mathcal{S}^k_2,
\ldots, \mathcal{S}^k_{c_k})$.  The state of the system can only ever change in one of two ways, either via
the infection of a node or via the recovery of a node.  We can describe the evolution in the state space by a 
continuous time Markov-process.  Setting $X^k_j(t)$ as the probability of the system being in state {$\mathcal{S}_j^k$}
at time $t$ and letting $X^k(t) = (X^k_2(t),X^k_2(t),\ldots,X^k_{c_k}(t)$ we can then write the Kolmogorov equations
that capture the two possible transitions in the following matrix and vector form,
\[
\dot{X}^k(t) = 
	\begin{cases} 
		B^0X^0 + C^0X^1 &\mbox{if } k = 0, \\
		A^kX^{k-1} + B^kX^k + C^kX^{k+1} &\mbox{for } 1 \leq k \leq N-1, \\
		A^NX^{N-1} + B^NX^N &\mbox{if } k = N
	\end{cases}
\]
Here the matrices $A^k$ capture the transitions into $\mathcal{S}^k$ via infection, $C^k$ capture the transitions into 
$\mathcal{S}^k$ via recovery and $B^k$ captures transitions within $\mathcal{S}^k$.  Their entries are given as
follows:

\begin{itemize}
\item		$A^k_{i,j}$ is the rate of transition from $\mathcal{S}_j^{k-1}$ to $\mathcal{S}_i^{k}$, where $k = 0,1,\ldots, N$,
			$i = 1,2,\ldots,c_k$ and $j = 1,2,\ldots,c_{k-1}$ .  Note that none-zero entries of the matrix represent
			the transitions where only one individual is changing from susceptible to infected and the corresponding 
			entrance will then equal $\tau$ multiplied by the number of infectious neighbours of the susceptible.  These
			matrices encode the topological structure of the network.
\item 		$C^k_{i,j}$ is the rate of transition from $\mathcal{S}_j^{k+1}$ to $\mathcal{S}_i^{k}$, where $k = 0,1,\ldots, N$,
			$i =1,2,\ldots,c_k$ and $j = 1,2,\ldots,c_{k+1}$.  Note that none-zero entries of the matrix represent
			the transitions where only one individual is changing from infected to susceptible and the corresponding 
			entrance will then equal $\gamma$.
\item		$B^k_{i,j}$ is the rate of transition from $\mathcal{S}_j^{k}$ to $\mathcal{S}_i^{k}$ where $B_{i,j}^k = 0$ if $i \neq j$ with 
			$k = 0,1,\ldots, N$ and $i,j = 1,2,\ldots,c_k$.
\end{itemize}

Letting $X(t) = (X^0(t),X^1(t),\ldots, X^N(t))^T$, we then write Kolmogorov equations in the following block tridiagonal
form, $\dot{X} = PX$, where
\begin{align*}
	P  &= \left(\begin{array}{cccccc}
				B^0 & C^0 &  0   &  0   &  0   & 0\\
				A^1 & B^1 & C^1  &  0   &  0   & 0\\
				 0  & A^2 & B^2  & C^2  &  0   & 0\\
				 0  &  0  & A^3  & B^3  & C^3  & 0\\
				 0  &  0  &\dots &\dots &\dots & 0\\
				 0  &  0  &  0   &  0   & A^N  & B^N\\
			\end{array}\right).
\end{align*}

From \cite{istvan1}, we also know that the entries of the matrix $B$ are zero except on the diagonals, where we find that
\begin{align}
B_{jj}^k &= -\sum_{i=1}^{c_{k+1}}{A_{i,j}^{k+1}} - \sum_{i=1}^{c_{k-1}}{C_{i,j}^{k-1}} \notag \\
				 &= -\tau N_{SI}(S_j^k) - k\gamma. \label{bjj}
\end{align}
Where \cite{istvan1} focussed on individual and edge motifs here we focus on the derivation of evolution equations
for the expected number of an arbitrary motif, $\hat{m}$.  We begin by writing the exact equations for an arbitrary 
motif $\hat{m}$ based on the transition and recovery matrices.  Using the notation from Table~{\ref{table:2}} this yields,
\begin{align}
\dot{|\mathcal{M}|}	= &\sum_{k=0}^N{N_{\hat{m}}(S^k)\dot{X}^k} \notag \\  
											= &N_{\hat{m}}(S^0)\left[B^0X^0 + C^0X^1\right] \notag \\
											  &+ \sum_{k=1}^{N-1}N_{\hat{m}}(S^k)\left[A^kX^{k-1} + B^kX^k + C^kX^{k+1}\right] \notag \\
											  &+ N_{\hat{m}}(S^N)\left[A^NX^{N-1} + B^NX^N\right] \notag \\
											= &\sum_{k=1}^{N}N_{\hat{m}}(S^k)A^kX^{k-1} + \sum_{k=0}^{N}N_{\hat{m}}(S^k)B^kX^k + \sum_{k=0}^{N-1}N_{\hat{m}}(S^k)C^kX^{k+1} \notag \\
											= &\sum_{k=0}^{N-1}N_{\hat{m}}(S^{k+1})A^{k+1}X^{k} + \sum_{k=0}^{N}N_{\hat{m}}(S^k)B^kX^k + \sum_{k=1}^{N}N_{\hat{m}}(S^{k-1})C^{k-1}X^{k} \notag \\
											= &\left[N_{\hat{m}}(S^1)A^1 + N_{\hat{m}}(S^0)B^0\right]X^0 \notag \\
											  &+ \sum_{k=1}^{N-1}\left[N_{\hat{m}}(S^{k+1})A^{k+1} + N_{\hat{m}}(S^k)B^{k} + N_{\hat{m}}(S^{k-1})C^{k-1}\right]X^k\notag\\
											  &+ \left[N_{\hat{m}}(S^N)B^N + N_{\hat{m}}(S^{N-1})C^{N-1}\right]X^N. 
\end{align}
Before continuing we note the following
\begin{align*}
B^N &= B_{1,1}^N = - \sum_{i=1}^{N}C_{i,1}^{N-1} = -\gamma N, \\
B^0 &= B_{1,1}^0 = - \sum_{i=1}^{N}A_{i,1}^{1} = -\tau N_{SI}(S_1^0) = 0. \\
\end{align*}
Taking these and (\ref{bjj}) into account and using the fact that $B$ is only none zero on it's diagonal, 
we then obtain the following equation,
\begin{align}
\dot{|\mathcal{M}|}	= &N_{\hat{m}}(S^1)A^1X^0 \nonumber \\
					&+ \sum_{k=1}^{N-1}\left[N_{\hat{m}}(S^{k+1})A^{k+1} -\tau\left(N_{\hat{m}}(S^k)*N_{SI}(S^k)\right) - \gamma kN_{\hat{m}}(S^k) \right.
					 + N_{\hat{m}}(S^{k-1})C^{k-1}\left.\right]X^k \nonumber \\
					&+ \left[N_{\hat{m}}(S^{N-1})C^{N-1} - \gamma NN_{\hat{m}}(S^N)\right]X^N \nonumber \\
				= &\sum_{k=1}^{N-1}\left[N_{\hat{m}}(S^{k+1})A^{k+1} -\tau\left(N_{\hat{m}}(S^k)*N_{SI}(S^k)\right)\right]X^k 
				 - \sum_{k=1}^{N}\left[\gamma kN_{\hat{m}}(S^k) - N_{\hat{m}}(S^{k-1})C^{k-1}\right]X^k.
\end{align}
We note that the term containing $X^0$ vanishes because $A^1$ is a column vector with all zero entries.
We now consider the summations involving the $A$ and $C$ matrices and the state $\mathcal{S}_j^k$.
In this state there are $k$ infected and $N-k$ susceptible individuals.  Without loss of generality the
susceptible individuals are numbered $1$ to $N-k$ and the infected numbered from $N-k+1$ to $N$.
Defining $r_t$ to be the number of infective neighbours of the node numbered $t$ we then obtain:
\begin{align*}
\left[N_{\hat{m}}(S^{k+1})A^{k+1}\right]_{j}
		= &\sum_{i=1}^{c_{k+1}}N_{\hat{m}}(S_i^{k+1})A_{i,j}^{k+1} \\\
		= &r_1\tau\left[N_{\hat{m}}(S_j^k) + (\text{number of } \hat{m} \text{ gained by node } 1 \text{ becoming infected })\right.\\
		&\left. - (\text{number of } \hat{m} \text{ lost by node } 1 \text{ becoming infected })		\right] \\ 	
		+ &r_2\tau\left[N_{\hat{m}}(S_j^k) + (\text{number of } \hat{m} \text{ gained by node } 2 \text{ becoming infected })\right.\\
		&\left. - (\text{number of } \hat{m} \text{ lost by node } 2 \text{ becoming infected })		\right] \\
		+ &\ldots \\
		+ &r_{N-k}\tau\left[N_{\hat{m}}(S_j^k) + (\text{number of } \hat{m} \text{ gained by node } (N-k) \text{ becoming infected })\right.\\
		&\left. - (\text{number of } \hat{m} \text{ lost by node } (N-k) \text{ becoming infected })		\right]\\
		= &r_1\tau\left[N_{\hat{m}}(S_j^k) + (\text{number of elements of } M_{k,j}^- \text{ where node } 1 \text{ is susceptible }\right.\\
									&\quad\quad\quad\quad\quad\text{and where node } 1's \text{ infection would lead to a motif of type } \hat{m})\\
		&\left. - (\text{number of elements of  } M_{k,j} \text{ where node } 1 \text{ is susceptible })		\right] \\ 	
		+ &r_2\tau\left[N_{\hat{m}}(S_j^k) + (\text{number of elements of } M_{k,j}^- \text{ where node } 2 \text{ is susceptible }\right.\\
									&\quad\quad\quad\quad\quad\text{and where node } 2's \text{ infection would lead to a motif of type } \hat{m})\\
		&\left. - (\text{number of elements of  } M_{k,j} \text{ where node } 2 \text{ is susceptible })		\right] \\ 	
		+ &\ldots \\
		+ &r_{N-k}\tau\left[N_{\hat{m}}(S_j^k) + (\text{number of elements of } M_{k,j}^- \text{ where node } (N-k) \text{ is susceptible }\right.\\
									&\quad\quad\quad\quad\quad\text{and where node } (N-k)'s \text{ infection would lead to a motif of type } \hat{m})\\
		&\left. - (\text{number of elements of  } M_{k,j} \text{ where node } (N-k) \text{ is susceptible })		\right], \\
\end{align*}
grouping the terms we obtain,
\begin{align*}
\left[N_{\hat{m}}(S^{k+1})A^{k+1}\right]_{j}
		= &\tau N_{SI}(S_j^k)N_{\hat{m}}(S_j^k) + 
		      \tau \sum_{i=1}^{|M_{k,j}^-|}\left[ N_{in}^{SI}(\hat{m}_{k,j}^{i-},\hat{m}) + N_{ex}^{SI}(\hat{m}_{k,j}^{i-},\hat{m})\right] \\
		     -&\tau|M_{k,j}|N_{in}^{SI}(\hat{m}) - \tau \sum_{i=1}^{|M_{k,j}|}\left[N_{ex}^{SI}(\hat{m}_{k,j}^i)\right].
\end{align*}
Similarly,
\begin{align*}
\left[N_{\hat{m}}(S^{k-1})C^{k-1}\right]_{j}
		= &\sum_{i=1}^{c_{k-1}}N_{\hat{m}}(S_i^{k-1})C_{i,j}^{k-1} \\\
		= &\gamma\left[N_{\hat{m}}(S_j^k) + (\text{number of } \hat{m} \text{ gained by node } (N-k+1) \text{ recovering })\right.\\
		&\left. - (\text{number of } \hat{m} \text{ lost by node } (N-k+1) \text{ recovering })		\right] \\ 	
		+ &\gamma\left[N_{\hat{m}}(S_j^k) + (\text{number of } \hat{m} \text{ gained by node } (N-k+2) \text{ recovering })\right.\\
		&\left. - (\text{number of } \hat{m} \text{ lost by node } (N-k+2) \text{ recovering })		\right] \\
		+ &\ldots \\
		+ &\gamma\left[N_{\hat{m}}(S_j^k) + (\text{number of } \hat{m} \text{ gained by node } (N) \text{ recovering })\right.\\
		&\left. - (\text{number of } \hat{m} \text{ lost by node } (N) \text{ recovering })		\right] \\
		= &\gamma\left[N_{\hat{m}}(S_j^k) + (\text{number of elements of  } M_{k,j}^+ \text{ where node } (N-k+1) \text{ is infective }\right. \\
									&\quad\quad\quad\quad\quad\text{and where node } (N-k+1)'s \text{ recovery would lead to a motif of type } \hat{m})\\
		&\left. - (\text{number of elements of } M_{k,j} \text{ of which node } (N-k+1) \text{ belongs })		\right] \\ 	
		+ &\gamma\left[N_{\hat{m}}(S_j^k) + (\text{number of elements of  } M_{k,j}^+ \text{ where node } (N-k+2) \text{ is infective }\right.\\
									&\quad\quad\quad\quad\quad\text{and where node } (N-k+2)'s \text{ recovery lead to a motif of type } \hat{m})\\
		&\left. - (\text{number of elements of } M_{k,j} \text{ of which node } (N-k+2) \text{ belongs }) \right] \\
		+ &\ldots \\
		+ &\gamma\left[N_{\hat{m}}(S_j^k) + (\text{number of elements of  } M_{k,j}^+ \text{ where node } (N) \text{ is infective }\right.\\
									&\quad\quad\quad\quad\quad\text{and where node } N's \text{ recovery would lead to a motif of type } \hat{m})\\
		&\left. - (\text{number of elements of } M_{k,j} \text{ of which node } (N) \text{ belongs }) \right],\\
\end{align*}
grouping the terms we obtain
\begin{align*}
\left[N_{\hat{m}}(S^{k-1})C^{k-1}\right]_{j}
		= &\gamma k N_{\hat{m}}(S_j^k) + \gamma \sum_{i=1}^{|M_{k,j}^+|} N^{I}(\hat{m}_{k,j}^{i+},\hat{m}) -\gamma|M_{k,j}|\left(N^I(\hat{m})\right).
\end{align*}
Defining
\begin{align*} 
\mathcal{A}^{k+1}_{j} &= \tau \sum_{i=1}^{|M_{k,j}^-|}\left[ N_{in}^{SI}(\hat{m}_{k,j}^{i-},\hat{m}) + N_{ex}^{SI}(\hat{m}_{k,j}^{i-},\hat{m})\right]
		     -\tau|M_{k,j}|N_{in}^{SI}(\hat{m}) - \tau \sum_{i=1}^{|M_{k,j}|}\left[N_{ex}^{SI}(\hat{m}_{k,j}^i)\right] \\
\mathcal{C}^{k-1}_{j} &= \gamma \sum_{i=1}^{|M_{k,j}^+|} \left[N^{I}(\hat{m}_{k,j}^{i+},\hat{m})\right] -\gamma|M_{k,j}|\left(N^I(\hat{m})\right)
\end{align*}
and setting $\mathcal{A}^{k+1} = [{A}^{k+1}_{1}, {A}^{k+1}_{j}, \ldots, {A}^{k+1}_{c_{k}}]$ and
$\mathcal{C}^{k-1} = [{C}^{k-1}_{1}, {C}^{k-1}_{j}, \ldots, {C}^{k-1}_{c_{k-1}}]$ yields,
\begin{align*}
\dot{|\mathcal{M}|}	= &\sum_{k=1}^{N-1}\left[N_{\hat{m}}(S^{k+1})A^{k+1} -\tau\left(N_{\hat{m}}(S^K)*N_{SI}(S^k)\right)\right]X^k 
				 	- \sum_{k=1}^{N}\left[\gamma kN_{\hat{m}}(S^k) - N_{\hat{m}}(S^{k-1})C^{k-1}\right]X^k \\
				 	= &\sum_{k=1}^{N-1}\left[\tau\left(N_{\hat{m}}(S^K)*N_{SI}(S^k)\right) + \mathcal{A}^{k+1}-\tau\left(N_{\hat{m}}(S^K)*N_{SI}(S^k)\right)\right]X^k -\\
					   &\sum_{k=1}^{N}\left[ \gamma kN_{\hat{m}}(S^k) - \left(kN_{\hat{m}}(S^k) + \mathcal{C}^{k-1} \right)\right]X^k \\
					= &\sum_{k=1}^{N-1}\left[\mathcal{A}^{k+1}\right]X^k + \sum_{k=1}^{N}\left[ \mathcal{C}^{k-1} \right]X^k \\
					= &\sum_{k=1}^{N-1}\sum_{j=1}^{c_k} \mathcal{A}_j^{k+1}X_j^k  + \sum_{k=1}^{N}\sum_{j=1}^{c_k}C_{j}^{k-1}X_{j}^{k} \\
					= &\sum_{k=1}^{N-1}\sum_{j=1}^{c_k}\left\{ \tau \sum_{i=1}^{|M_{k,j}^-|}\left[ N_{in}^{SI}(\hat{m}_{k,j}^{i-},\hat{m}) + N_{ex}^{SI}(\hat{m}_{k,j}^{i-},\hat{m})\right] -\tau|M_{k,j}|N_{in}^{SI}(\hat{m}) - \tau \sum_{i=1}^{|M_{k,j}|}\left[N_{ex}^{SI}(\hat{m}_{k,j}^i)\right]\right\}X_{j}^k \\
					   &+\sum_{k=1}^{N}\sum_{j=1}^{c_k}\left\{\gamma \sum_{i=1}^{|M_{k,j}^+|}\left[ N^{I}(\hat{m}_{k,j}^{i+},\hat{m})\right] -\gamma|M_{k,j}|\left(N^I(\hat{m})\right)\right\}X_{j}^{k}\\
					= &\tau\mathcal{N}_{in}^{SI}(\hat{m}^-,\hat{m}) + \tau\mathcal{N}_{ex}^{SI}(\hat{m}^-,\hat{m}) -\tau|\mathcal{M}|N_{in}^{SI}(\hat{m}) - \tau\mathcal{N}_{ex}^{SI}(\hat{m}) \notag \\
					&+\gamma\mathcal{N}^{I}(\hat{m}^+,\hat{m}) - \gamma|\mathcal{M}|N^{I}(\hat{m}).
\end{align*}
Which matches equation~\ref{heuristic} from Theorem~1. It is worth noting that our result is
related to the equation for the ``expectation of some average quantity'' given in~{\cite{rand1}}.  However,
whilst the result in~{\cite{rand1}} is very general here we provide a proof by construction that, for a given 
motif, pinpoints the events that influence these motif and their rates.
\label{sec:5a}

\subsection{Using the Theorem to prove the conjectured exact effective degree model is derivable from the Kolmogorov equations}
Letting $\hat{m}$ be an $S_{s,i}$-type motif from the effective degree model earlier and using Theorem~$1$, we find that 
the exact equations can be written as
\begin{align*}
\frac{d S_{s,i}}{dt} = 	&\tau\mathcal{N}_{in}^{SI}(\hat{m}^-,\hat{m}) + \tau\mathcal{N}_{ex}^{SI}(\hat{m}^-,\hat{m}) -\tau|\mathcal{M}|N_{in}^{SI}(\hat{m}) - \tau\mathcal{N}_{ex}^{SI}(\hat{m}) \notag \\
					&+\gamma\mathcal{N}^{I}(\hat{m}^+,\hat{m}) - \gamma|\mathcal{M}|N^{I}(\hat{m}) \\
				    = &\tau\times\left(\text{the total expected number of SI connections within } S_{s+1,i-1} \text{-type motifs}\right.\\
				    							&\quad\quad\text{where if infection occurs we obtain a } S_{s,i} \text{-type motif}\left.\right) \\
				       &+ \tau\times\left(\text{the total expected number of SI connections where S lies within} \right.\\
				       &\quad\quad\quad S_{s+1,i-1} \text{-type motifs and the I is external to the given motif}\\
				       &\quad\quad\quad \text{and where, were an infection to occur, we obtain a } S_{s,i} \text{-type motif}\left.\right) \\
				       &- \tau S_{s,i}\times\left(\text{number of SI connections within an individual } S_{s,i}\text{-type motifs}\right) \\
				       &- \tau \times\left(\text{the total expected number of SI connections where S belongs to } \right. \\
				       &\quad\quad\quad S_{s,i}\text{-type motifs and the I is external to the given motif} \left.\right)\\
				       &+\gamma\times\left(\text{the total expected number I's within } S_{s-1,i+1} \text{-type and } I_{s,i} \text{-type motifs}\right. \\
				       &\quad\quad\quad \text{where there recovery would give a } S_{s,i} \text{-type motif}\left.\right) \\
				       &-\gamma S_{s,i}\times\left(\text{number of I within an individual } S_{s,i} \text{-type motif}\right) \\
			          = &\tau\left[ISS_{s+1,i-1}\right] -\tau i S_{s,i} - \tau\left[ISS_{s,i}\right] + \gamma I_{si} 
		  							 	 +\gamma (i+1)S_{s-1,i+1}-\gamma iS_{s,i}
\end{align*}
which is indeed the conjectured exact equation for $S_{s,i}$ (similar derivation holds for $I_{s,i}$).  To clarify the above derivation
we note that a term such as $\tau\mathcal{N}_{in}^{SI}(\hat{m}^-,\hat{m})$ will make no contribution to the resultant equation
as there are no internal $SI$ connections within $S_{s-1,i+1}$-type motifs along which an infection would lead to an $S_{s,i}$-type motif.
However other terms, such as $\tau\mathcal{N}_{ex}^{SI}(\hat{m}^-,\hat{m})$, have a direct correspondence with the resultant output
(in this case the $\tau\left[ISS_{s+1,i-1}\right]$ term).
\label{sec:5b}

\section{Comparison of the closed models}
In comparing the models the obvious question to ask is when does one model perform better than another, i.e.\ which model
approximates better or more accurately the simulation results or the solution of the Kolmogorov/master equations where
solvable.  As discussed earlier, the pairwise model is known to perform well on networks that are well characterised by the 
average degree (i.e.\ regular random and Erd\H{o}s-R\'enyi graphs).  What is less known is under what circumstances do the 
heterogenous pairwise and effective degree models outperform one another.

\begin{figure}[htp!]
    \begin{center}
        \subfigure[Regular random]{
            \label{fig3a}
            \includegraphics[width=0.75\textwidth]{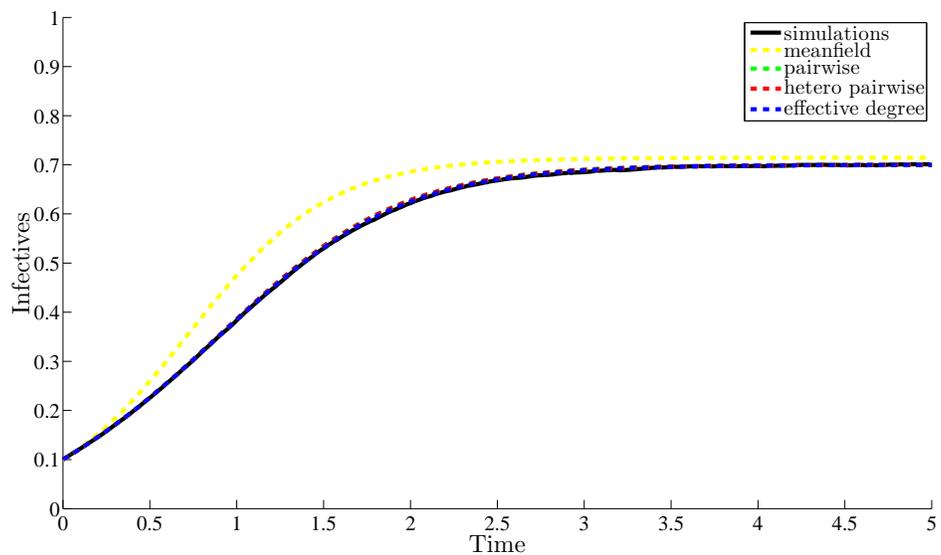}
        }\\
        \subfigure[Erdos-Renyi]{
           \label{fig3b}
           \includegraphics[width=0.75\textwidth]{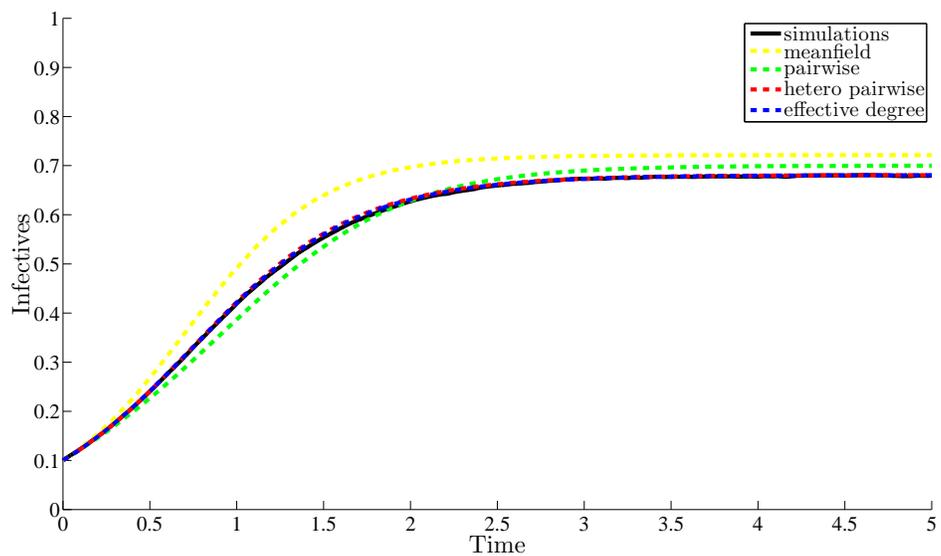}
        }       
    \end{center}
    \caption{{\bf ODE performance on different networks} Each network is of size $N=500$ and with disease
    			parameters given by $\gamma = 1$ and $\tau = 0.5$.  Average prevalence was calculated from individual simulations
    			on $500$ different networks.  Initial conditions for the ODEs were obtained by averaging the initial
						conditions from each of the simulations. (a) Regular random network, each node having degree $7$.  (b) 
			Erd\H{o}s-R\'enyi random network with average degree $7$.}
    \label{fig3}
\end{figure}

To assess the performance of the three closed models we compared individual simulations to the solutions of 
the ODE's on four different types of undirected network.  For each of the different types of networks we used the
Gillespie algorithm, {\cite{gillespie}}, to run $500$ simulations on networks of size $N=500$  ($1$ simulation on $500$ different randomly generated networks).
  The results of these simulations were then averaged to obtain an expected value to compare to the solution of the various ODE's.  We began by considering regular random networks 
where all nodes have the same number of randomly chosen neighbours  and then Erd\H{o}s-R\'enyi random networks
where the distribution of degrees converges to a Poisson distribution.  Figure~\ref{fig3} plots simulation 
results against the different solutions of the ODEs for these two networks.  On the regular network, 
whilst the two different pairwise models and the effective degree offer an improvement in performance over the standard 
meanfield equations, there is little to distinguish between the improved approaches.  As expected, on the Erd\H{o}s-R\'enyi
random networks, the pairwise model improves on the meanfield model and, in turn, the effective 
degree and heterogeneous pairwise models improve even further on this.  Again, however, there is little 
to distinguish between effective degree and the heterogeneous pairwise models.  

\begin{figure}[htp!]
    \begin{center}
        \subfigure[Power-law]{
            \label{fig4a}
            \includegraphics[width=0.75\textwidth]{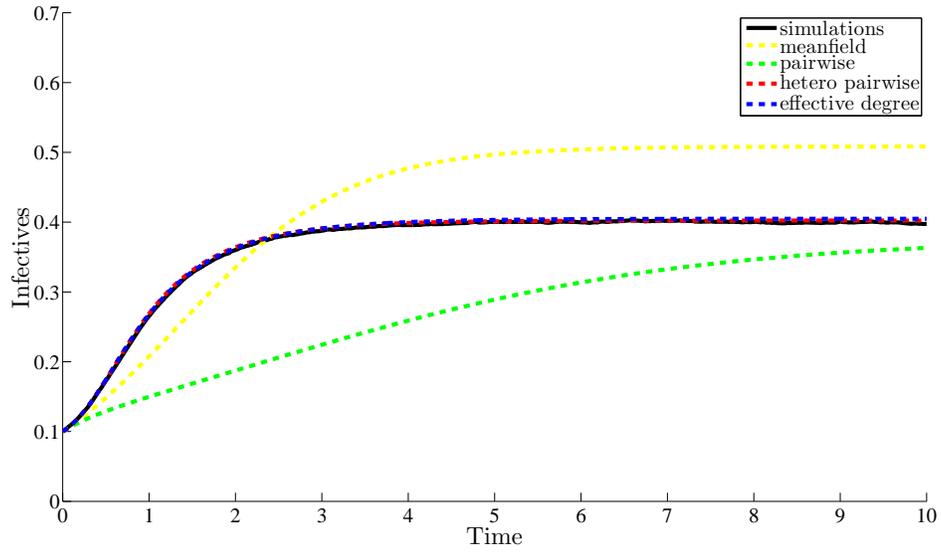}
        }\\
        \subfigure[Assortative power-law]{
           \label{fig4b}
           \includegraphics[width=0.75\textwidth]{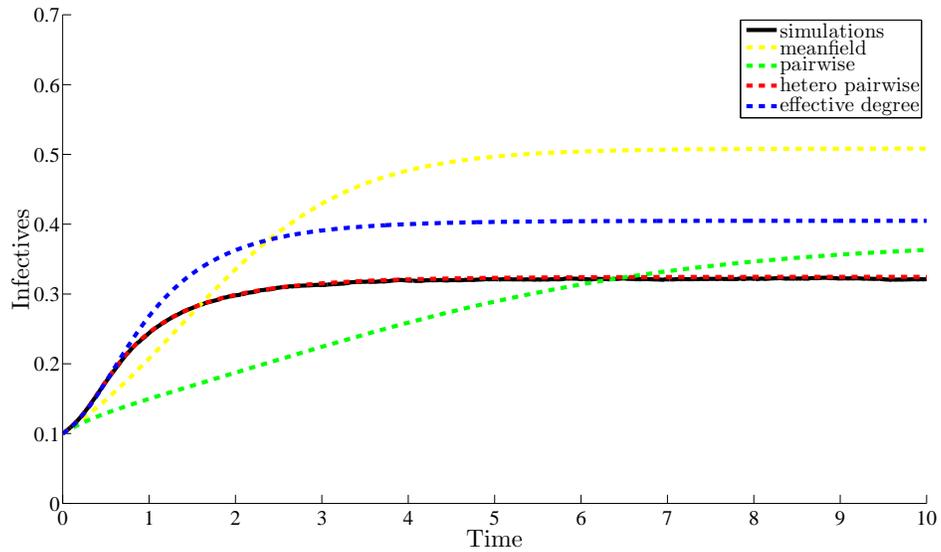}
        }       
    \end{center}
    \caption{{\bf ODE performance on different networks.} Each network is of size $N=500$ and with disease
    			parameters given by $\gamma = 1$ and $\tau = 0.5$. Average prevalence was calculated from individual simulations
    			on $500$ different networks.  Initial conditions for the ODEs were obtained by averaging the initial
						conditions from each of the simulations.   (a) Network with degrees chosen from a power law distribution  (b) 
						Networks with degrees chosen from a powerlaw distribution but rewired to have assortativity coefficient
						$r \approx 0.49$.}
    \label{fig4}
\end{figure}

To investigate further we ran simulations on networks exhibiting greater heterogeneity in their degree distribution.  Firstly 
we considered networks with degrees between $1$ and $25$ chosen from a powerlaw degree distribution ($p(x) = Ax^{-1.5}$) and
generated by the configuration model algorithm {\cite{newmanconfig}}.  Networks with scale-free like degree distributions 
may be more closely related to those of real world networks ({\cite{barabasi}}) and may thus be of greater use in understanding the
applicability of more theoretical modelling approaches.   Secondly we considered graphs with the same power law
degree distributions as before but this time rewired based on the ``greedy" assortativity algorithm (discussed in {\cite{winterbach}}).
This rewiring leads to an increase in the assortativity coefficient ({\cite{newman1}}) which measures the propensity of nodes of similar
degrees to attach to one another.  In theory, we should be able to capture this correlation  with the heterogenous
pairwise equations as they explicitly take the degree of connected nodes into account within the initial conditions.
The results are illustrated in Figure~{\ref{fig4}}.

Whilst on the powerlaw network network there is little difference between heterogeneous pairwise and effective degree when the 
assortativity is increased, there is a clear improvement in the performance of the heterogenous pairwise model over the effective 
degree.  Any performance benefit must, however, be considered in terms of the model complexity given in table~{\ref{table:3}} 
(note that in this table $M$ is the maximum possible degree in the network and we given the minimum number of equations needed to 
implement the ODEs).

\begin{table}[htp!]
\caption{Complexity of closed ODEs}
\label{table:3}      
\begin{tabular}{lll}
\hline\noalign{\smallskip}
Model & \# equations & complexity  \\
\noalign{\smallskip}\hline\noalign{\smallskip}
meanfield & $1$ & $\mathcal{O}(1)$ \\
pairwise & $3$ & $\mathcal{O}(1)$ \\
effective degree & $M(M+3)-1$ & $\mathcal{O}(M^2)$ \\
heterogeneous pairwise & $2M(M+1)-1$ & $\mathcal{O}(M^2)$ \\
Kolmogorov equations & $2^N$ & $\mathcal{O}(2^N)$ \\
\noalign{\smallskip}\hline
\end{tabular}
\end{table}

\begin{figure}[htp!]
  \begin{center}
    \includegraphics[width=0.75\textwidth]{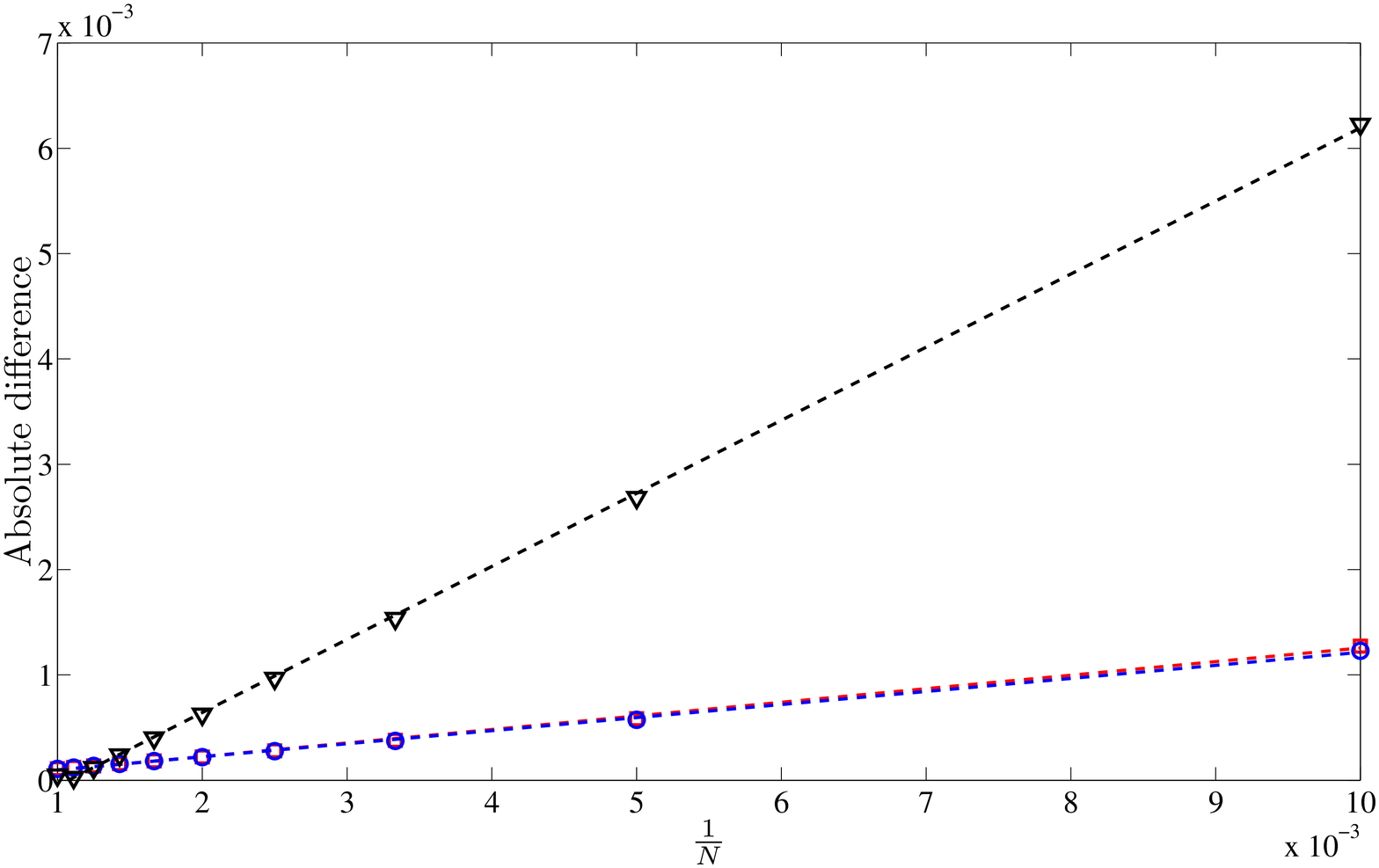}
  \end{center}
    \caption{{\bf Convergence to exact solution on a complete graph.}  Absolute difference between
    the exact steady state solution of the percentage of infected individuals and those calculated from three
    different ODE models for $10$ different network sizes and initial prevalence of $40$ percent.  
    Black triangles represent meanfield, blue circles effective 
    degree and red squares the pairwise equations. Linear lines of best fit are also shown.
    This shows that the error(N) appears to be of O(1/N) as N tends to infinity.} 
  \label{fig5}
\end{figure}

A final comparison between the performance of the different closed models is to look at their rate of convergence to the solution of
the Kolmogorov equations on a complete (fully connected) network.  On a complete network it is possible (see \cite{istvan1}) to reduce the 
full system of $2^N$ equations to just $N+1$ equations.  This allows us to compare the true solution to the approximate solution of 
the meanfield, pairwise (equivalent to heterogenous pairwise on a complete graph) and effective degree models.  Interestingly we find that
all three exhibit $O(1/N)$ convergence, where although both pairwise and effective degree bring an improvement on meanfield, 
the difference between the convergence of the two is neglible and almost indecernible (see figure~\ref{fig5}).
\label{sec:6}

\section{Discussion}
\begin{figure}[htp!]
  \begin{center}
    \includegraphics[width=0.5\textwidth]{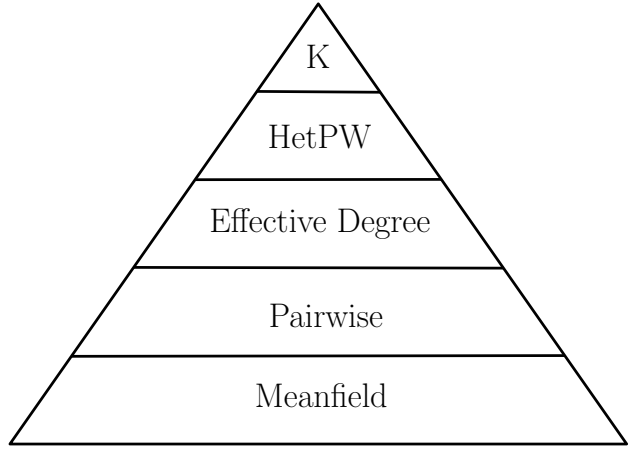}
  \end{center}
    \caption{{\bf Model performance hierarchy.}  Model performance hierarchy based on our observations.
    				Here K represents the Kolmogorov equations and HetPW the heterogeneous pairwise equations.}
  \label{fig6}
\end{figure}

In this paper we set out to achieve a greater understanding of the relation between some of the more common approaches
to modelling disease dynamics.  In doing so we conjectured an exact version of the effective degree model~\cite{lindquist} 
and showed how this model could be used to recover the pairwise model \cite{keeling1}.  We then extended this model to 
incorporate greater network structure and illustrated how, from this extension, we could then recover the heterogeneous 
pairwise model \cite{eames1}.  We then proved that the conjectured exact effective degree model was indeed exact by proving
that a heuristic derivation of an ODE model for an arbitrary motif was derivable directly from the Kolmogorov equations
and noting that the exact effective degree model was just a particular case of this heuristic model.  Finally we considered the 
performance of the different models on four different type of networks and have analysed numerically the rate of convergence
to the lumped Kolmogorov equations on a complete network.  These comparisons suggest a performance hierarchy of models
as illustrated in Figure~\ref{fig6} and it is worth noting that the performance benefit of the heterogenous pairwise model 
on networks exhibiting susceptible $\rightarrow$ infectious $\rightarrow$ removed (SIR) disease dynamics was also 
touched upon in \cite{danon}.

Whilst we have shown how current models can be extended in a way that can capture more network topology, these extensions
have a more theoretical rather than practical motivation as their added complexity makes them not only less tractable but
also more resource intensive in their solving, thus making the use of simulations more of an attractive proposition.
As the links between these models are better understood, future work will likely focus on the following three areas.  Firstly,
a more realistic network will have a more clique-like structure.  For example an individual is likely a member of a household 
in which he has regular contacts within and less regular contacts outside.  Being able to incorporate this household structure
within epidemic models is thus important in understanding the outbreak and necessary curtailment of an infectious disease 
(see \cite{ball,house,volz}).
Secondly, a network of individuals is not well represented by a static network.  An individual may have regular contact with few
individuals but may create or break contacts with others in ways that a static network representation cannot capture.  
For this reason it is important to take into consideration not only the dynamics of the disease but also the dynamics of the network and 
how the two impact on one another (see~\cite{gross,istvan2}).  Thirdly, assuming we can write down exact differential equations we have to close them in
some way.  Understanding the performance of current, and also the derivation of new closures, is arguably the most important task 
ahead as it is the closures that limit the performance of any system of ODEs.

\label{sec:7}

\newpage
\appendix
\section*{Appendix $1$}
\label{Appendix1}
Illustration of the exact effective degree transitions where the central node is infective.
\begin{figure}[htp!]
  	\begin{center}
    		\includegraphics[width=0.75\textwidth]{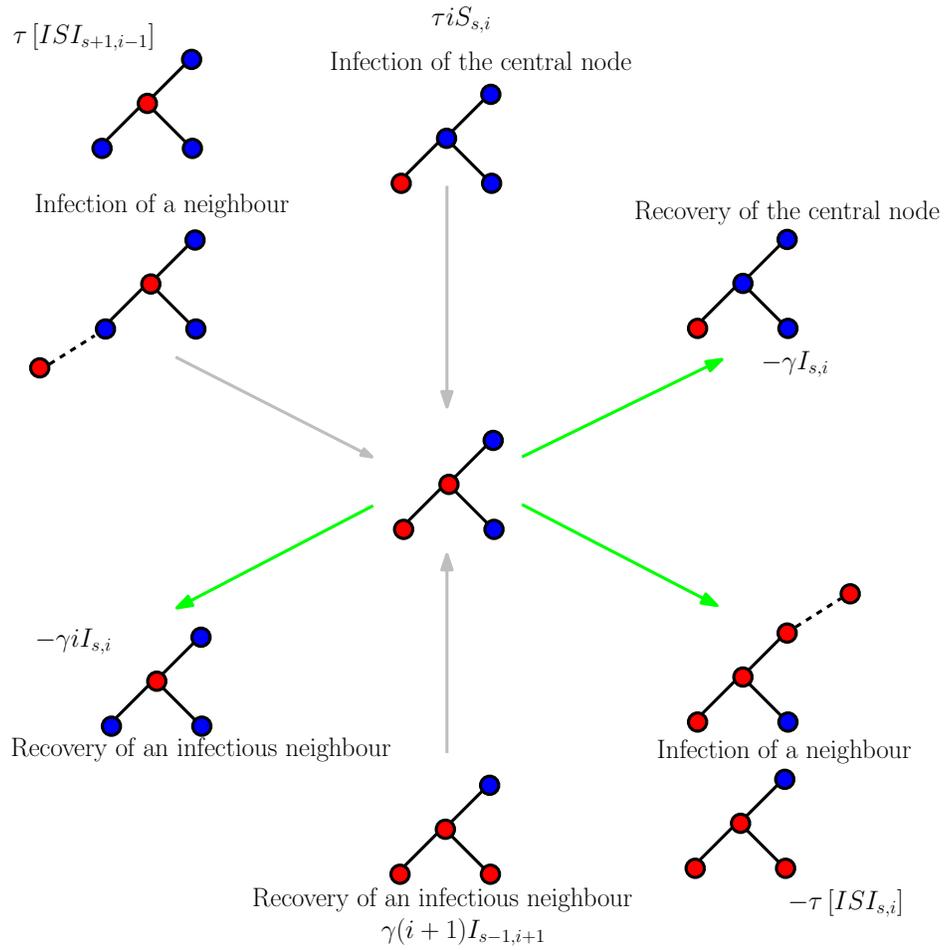}
  	\end{center}
  	\caption{Illustration of the transitions into and out of the $I_{2,1}$ class.  Susceptible nodes are given in blue
  		       	   and infective nodes in red.  Transitions into and out of the class are shown in grey and green, respectively.
  				   The corresponding terms of the general equation are also given.}  
	\label{fig7}
\end{figure}
\label{sec:appendix1}

\section*{Appendix $2$}
\label{Appendix2}
Derivation of the pairwise equation from the exact effective degree model for singles and pairs are as follows,
\begin{align*}
		\frac{d}{dt}\left[S\right] &= \sum_{s,i}\dot{S}_{s,i} = \gamma\left[I\right] - \tau\left[S I\right], \\
		\frac{d}{dt}\left[I\right] &= \sum_{s,i}\dot{I}_{s,i} = -\gamma\left[I\right] + \tau\left[S I\right],
\end{align*}
where most terms from the original effective degree equations cancel and we have used that $\sum_{s,i}iS_{s,i} = [SI]$ and
$\sum_{s,i}I_{s,i} = [I]$.  For the pairs the effective degree model yields,
\begin{align*}
\frac{d}{dt}\left[SS\right] 
			= &\sum_{s,i}s\dot{S}_{si} \\
			= &-\tau\sum{siS_{s,i}} + \gamma\sum{sI_{s,i}}
				 +\gamma\sum{s(i+1)S_{s-1,i+1}} -\gamma\sum{isS_{s,i}} \\
				&+\tau\sum{s[ISS_{s+1,i-1}]} - \tau\sum{s[ISS_{s,i}]} \\
			= &-\tau[ISS] + \gamma[IS] + \gamma\sum{(s-1)(i+1)S_{s-1,i+1}} + \gamma\sum{(i+1)S_{s-1,i+1}} \\
				&-\gamma[ISS] + \tau\sum{(s+1)[ISS_{s+1,i-1}]} - \tau\sum{[ISS_{s+1,i-1}]} - \tau\sum{s[ISS_{s,i}]} \\
			= &-\tau[ISS] + \gamma[IS] + \gamma[ISS] + \gamma[IS] -\gamma[ISS] -\tau[ISS] \\
			= &-2\tau[ISS] + 2\gamma[IS],
\end{align*}

\begin{align*}
\frac{d}{dt}\left[SI\right] 
			= &\sum_{s,i}s\dot{I}_{si} \\
			= &\tau\sum{siS_{s,i}} - \gamma\sum{sI_{s,i}}
				 +\gamma\sum{s(i+1)I_{s-1,i+1}} -\gamma\sum{siI_{s,i}} \\
				&+\tau\sum{s[ISI_{s+1,i-1}]} - \tau\sum{s[ISI_{s,i}]} \\
			= &\tau[ISS] - \gamma[IS] + \gamma\sum{(s-1)(i+1)I_{s-1,i+1}} + \gamma\sum{(i+1)I_{s-1,i+1}} \\
				&-\gamma[IIS] + \tau\sum{(s+1)[ISI_{s+1,i-1}]} - \tau\sum{[ISI_{s+1,i-1}]} - \tau\sum{s[ISI_{s,i}]} \\
			= &\tau[ISS] - \gamma[IS] + \gamma[IIS] + \gamma[II]-\gamma[IIS] -\tau([ISI]+[IS]) \\
			= &\tau\left([ISS] - [ISI] - [IS]\right) + \gamma\left([II] - [IS]\right),	
\end{align*}

\begin{align*}
\frac{d}{dt}\left[II\right] 
			= &\sum_{s,i}i\dot{I}_{si} \\
			= &\tau\sum{i^2S_{s,i}} - \gamma\sum{iI_{s,i}}
				 +\gamma\sum{i(i+1)I_{s-1,i+1}} -\gamma\sum{i^2I_{s,i}} \\
				&+\tau\sum{i[ISI_{s+1,i-1}]} - \tau\sum{i[ISI_{s,i}]} \\
			= &\tau\sum{i(i-1)S_{s,i}} + \tau\sum{iS_{s,i}} - \gamma[II] \\
				&+\gamma[III] - \gamma\sum{i(i-1)I_{s,i}} -\gamma\sum{iI_{s,i}} \\
				&+\tau\sum{(i-1)[ISI_{s+1,i-1}]} + \tau\sum{[ISI_{s+1,i-1}]} - \tau\sum{i[ISI_{s,i}]} \\
			= &\tau[ISI] + \tau[IS] - \gamma[II] + \gamma[III] -\gamma[III] - \gamma[II] + \tau[ISI] + \tau[IS] \\
			= &2\tau\left([ISI] + [IS]\right) - 2\gamma[II].
\end{align*}
\label{sec:appendix2}

\section*{Appendix $3$}
\label{Appendix3}
Derivation of the heterogeneous pairwise equations from the effective degree with neighbourhood composition model.  
For singles and pairs the following identities hold,

\begin{align*}
		\frac{d}{dt}\left[S^n\right] &= \sum_{|s'|+|i'|=N}\dot{S}_{s',i'} = \gamma\left[I^n\right] - \tau\left[S^n I\right], \\
		\frac{d}{dt}\left[I^n\right] &= \sum_{|s'|+|i'|=N}\dot{I}_{s',i'} =  -\gamma\left[I^n\right] + \tau\left[S^n I\right],
\end{align*}

\begin{align*}
\frac{d}{dt}\left[S^lS^n\right] 
			= &\sum_{|s'|+|i'|=n}{s_l\dot{S}_{s',i'}} \\
			= &-\tau\sum{s_l|i'|S_{s',i'}} + \gamma\sum{s_lI_{s',i'}}
				 +\gamma\sum{s_l\sum_{k=1}^{M}{(i_{k}+1)S_{s'_{k-},i'_{k+}}}} \\
			  &-\gamma\sum{s_l|i'|S_{s',i'}} + \tau\sum{s_l\sum_{k=1}^{M}{\left[IS^kS_{s'_{k+},i'_{k-}}\right]}
			   - \tau\sum{s_{l}\left[ISS_{s',i'}\right]}} \\
			= &-\tau\left[IS^nS^l\right] + \gamma\left[S^lI^n\right] + \gamma\sum{s_l\sum_{k \neq l}{(i_{k}+1)S_{s'_{k-},i'_{k+}}}} \\
			  &+\gamma\sum{(s_{l}-1)(i_{l}+1)S_{s'_{l-},i'_{l+}}} + \gamma\sum{(i_l+1)S_{s'_{l-},i'_{l+1}}} - \gamma\left[IS^nS^l\right] \\
			  &+ \tau\sum{s_l\sum_{k \neq l}{\left[IS^kS_{s'_{k+},i'_{k-}}\right]} + \tau\sum{(s_l+1)\left[IS^lS_{s'_{l+},i'_{l-}}\right]}} \\
			  &-\tau\sum{\left[IS^lS_{s'_{l+},i'_{l-}}\right] - \tau\sum{s_{l}\left[ISS_{s',i'}\right]}} \\
			= &-\tau\left[IS^nS^l\right] + \gamma\left[S^lI^n\right] + \gamma\left[S^lS^nI\right] + \gamma\left[I^lS^n\right] \\
			  &-\gamma\left[IS^nS^l\right]  - \tau\left[IS^lS^n\right]  \\
			= &-\tau\left[IS^nS^l\right] - \tau\left[IS^lS^n\right] + \gamma\left[S^lI^n\right] + \gamma\left[I^lS^n\right], \\
\end{align*}

\begin{align*}
\frac{d}{dt}\left[S^lI^n\right] 
			= &\sum_{|s'|+|i'|=n}{s_l\dot{I}_{s',i'}} \\
			= &\tau\sum{s_l|i'|S_{s',i'}} - \gamma\sum{s_lI_{s',i'}}
				 +\gamma\sum{s_l\sum_{k=1}^{M}{(i_{k}+1)I_{s'_{k-},i'_{k+}}}} \\
			  &-\gamma\sum{s_l|i'|I_{s',i'}} + \tau\sum{s_l\sum_{k=1}^{M}{\left[IS^kI_{s'_{k+},i'_{k-}}\right]}}
			   - \tau\sum{s_{l}\left[ISI_{s',i'}\right]} \\ 
			= &\tau\left[IS^nS^l\right] - \gamma\left[S^lI^n\right] + \gamma\sum{s_l\sum_{k \neq l}{(i_{k}+1)I_{s'_{k-},i'_{k+}}}} \\
			  &+\gamma\sum{(s_{l}-1)(i_{l}+1)I_{s'_{l-},i'_{l+}}} + \gamma\sum{(i_l+1)I_{s'_{l-},i'_{l+1}}} - \gamma\left[II^nS^l\right] \\
			  &+ \tau\sum{s_l\sum_{k \neq l}{\left[IS^kI_{s'_{k+},i'_{k-}}\right]}} + \tau\sum{(s_l+1)\left[IS^lI_{s'_{l+},i'_{l-}}\right]} \\
			  &-\tau\sum{\left[IS^lI_{s'_{l+},i'_{l-}}\right]} - \tau\sum{s_{l}\left[ISI_{s',i'}\right]}\\
			= &\tau\left[IS^nS^l\right] - \gamma\left[S^lI^n\right] + \gamma\left[S^lI^nI\right] + \gamma\left[I^lI^n\right] \\
			  &-\gamma\left[II^nS^l\right]  - \tau\left[IS^lI^n\right] - \tau\left[S^lI^n\right]  \\
			= &\tau\left[IS^nS^l\right] - \tau\left[IS^lI^n\right] - \tau\left[S^lI^n\right] + \gamma\left[I^lI^n\right] - \gamma\left[S^lI^n\right],   
\end{align*}  	

\begin{align*}
\frac{d}{dt}\left[I^lI^n\right] 
			= &\sum_{|s'|+|i'|=n}{i_l\dot{I}_{s',i'}} \\
			= &\tau\sum{i_l|i'|S_{s',i'}} - \gamma\sum{i_lI_{s',i'}}
				 +\gamma\sum{i_l\sum_{k=1}^{M}{(i_{k}+1)I_{s'_{k-},i'_{k+}}}} \\
			  &-\gamma\sum{i_l|i'|I_{s',i'}} + \tau\sum{i_l\sum_{k=1}^{M}{\left[IS^kI_{s'_{k+},i'_{k-}}\right]}}
			   - \tau\sum{i_{l}\left[ISI_{s',i'}\right]} \\
			= &\tau\sum{i_l\left(|i'|-1\right)S_{s',i'}} + \tau\sum{i_lS_{s',i'}} - \gamma\left[I^lI^n\right] \\
			  &+\gamma\left[I^lI^nI\right] -\gamma\sum{i_l\left(|i'|-1\right)I_{s',i'}} - \gamma\sum{i_lI_{s',i'}} \\
			  &+ \tau\sum{i_l\sum_{k \neq l}{\left[IS^kI_{s'_{k+},i'_{k-}}\right]}}
			  + \tau\sum{\left(i_l-1\right)\left[IS^lI_{s'_{l+},i'_{l-}}\right]}\\
			  &+ \tau\sum{\left[IS^lI_{s'_{l+},i'_{l-}}\right]} - \tau\sum{i_{l}\left[ISI_{s',i'}\right]}\\
			= &\tau\left[I^lS^nI\right] + \tau\left[I^lS^n\right] - 2\gamma\left[I^lI^n\right] \\
			  &+\gamma\left[I^lI^nI\right] - \gamma\left[I^lI^nI\right] + \tau\left[IS^lI^n\right] + \tau\left[S^lI^n\right] \\
			= &\tau\left[I^lS^nI\right] + \tau\left[I^lS^n\right] - 2\gamma\left[I^lI^n\right] + \tau\left[IS^lI^n\right] + \tau\left[S^lI^n\right].
\end{align*}
\label{sec:appendix3}


\end{document}